\documentclass[a4paper,aps,amsfonts,amsmath,nofootinbib,showpacs,byrevtex,showkeys,prd
]{revtex4}   

\usepackage{graphicx,bm}

\renewcommand{\vec}[1]{\ensuremath{\bm{\mathrm{#1}}}}%
\newcommand*{\uvec}[1]{\ensuremath{\Hat{\bm{\mathrm{#1}}}}}%

\renewcommand{\i}{\ensuremath{\mathrm{i}}}%

\newcommand*{\dbar}[1][]{\mathrm{d}\mkern-7mu\mathchar'26\mkern-1mu^{#1}}

\newcommand*{\be}{\begin{equation}}
\newcommand*{\ee}{\end{equation}}
\newcommand*{\braket}[2]{ \langle #1 \vert #2 \rangle }
\newcommand*{\bra}[1]{\langle #1 \rvert \mkern2mu}
\newcommand*{\ket}[1]{\mkern2mu \lvert #1 \rangle}
\newcommand*{\vev}[1]{\left< #1 \right>}

\DeclareMathOperator{\Tr}{Tr}			
\DeclareMathOperator{\Det}{Det}			

\newcommand*{\vx}{\vec{x}}
\newcommand*{\vy}{\vec{y}}
\newcommand*{\vp}{\vec{p}}
\newcommand*{\vq}{\vec{q}}
\newcommand*{\vk}{\vec{k}}
\newcommand*{\vl}{\vec{\ell}}
\newcommand*{\uvp}{\uvec{p}}
\newcommand*{\uvq}{\uvec{q}}
\newcommand*{\uvk}{\uvec{k}}

\renewcommand*{\d}[1][]{\mathrm{d}^{#1}} 
\newcommand*{\abs}[1]{\ensuremath{\lvert#1\rvert}}
\newcommand*{\eps}{\ensuremath{\varepsilon}}
\newcommand*{\e}{\ensuremath{\mathrm{e}}}
\newcommand*{\calD}{\ensuremath{\mathcal{D}}}
\newcommand*{\calJ}{\ensuremath{\mathcal{J}}}
\newcommand*{\calO}{\ensuremath{\mathcal{O}}}

\begin{document}

\title{Non-Gaussian wave functionals in Coulomb gauge Yang--Mills theory}

\author{Davide R.~Campagnari}
\author{Hugo Reinhardt}
\affiliation{Institut f\"ur Theoretische Physik, Universit\"at T\"ubingen,
Auf der Morgenstelle 14, 72076 T\"ubingen, Germany}
\date{\today}


\begin{abstract}
A general method to treat non-Gaussian vacuum wave functionals in the Hamiltonian
formulation of a quantum field theory is presented. By means of Dyson--Schwinger
techniques, the static Green functions are expressed in terms of the kernels arising
in the Taylor expansion of the exponent of the vacuum wave functional. These kernels
are then determined by minimizing the vacuum expectation value of the Hamiltonian.
The method is applied to Yang--Mills theory in Coulomb gauge, using a vacuum wave
functional whose exponent contains up to quartic terms in the gauge field. An estimate
of the cubic and quartic interaction kernels is given using as input the gluon and ghost
propagators found with a Gaussian wave functional.

\end{abstract}

\maketitle

\section{Introduction}

According to our present understanding of nature, Quantum Chromodynamics (QCD) is the
theory of the strong interaction. This theory has been tested in the high-momentum or
ultraviolet (UV) regime, where perturbation
theory is applicable due to asymptotic freedom. Our knowledge on the low energy, strongly
interacting regime of QCD stems mainly from lattice calculations, which have at least
qualitatively reproduced many physical observables, in particular, the linearly rising
confining potential for heavy quarks. Furthermore, lattice calculations have revealed
the relevance of topological field configurations such as magnetic monopoles and center
vortices for infrared phenomena like confinement and spontaneous chiral
symmetry breaking. These calculations support the dual Meissner effect and the vortex condensation picture of
confinement \cite{Gre03}. Despite these substantial physical insights provided by the
lattice calculations, a thorough understanding of these infrared phenomena will not
come from lattice calculations alone but will require also studies of the continuum
theory. 

In recent years there have been substantial efforts devoted to a non-perturbative
treatment of continuum Yang--Mills theory. Among these are a variational solution of
the Yang--Mills Schr\"odinger equation in Coulomb gauge \cite{SzcSwa01,Szc04,FeuRei04,ReiFeu05}.
In this approach, using Gaussian type wave functionals, minimization of the energy
density results in the so-called gap equation for the gluon energy (or static gluon
propagator). This equation has been solved analytically in the infrared \cite{SchLedRei06}
and in the ultraviolet \cite{CamReiWeb09} and numerically in the full momentum regime
\cite{FeuRei04,EppReiSch07}. One finds a gluon energy, which in the UV behaves
like the photon energy but diverges in the infrared (IR), signalling confinement. The obtained
gluon energy also compares favourably with the lattice data \cite{BurQuaRei09}. In particular,
the infrared regime is correctly reproduced, as far as we can tell from available lattice data.
There are, however, deviations in the mid-momentum regime
(and minor ones in the UV) which can be attributed to the
missing gluon loop, which escapes the Gaussian wave functionals. These deviations are
presumably irrelevant for the confinement properties, which are dominated by the ghost
loop (which is fully included under the Gaussian ansatz), but are believed to be important
for a correct description of spontaneous breaking of chiral
symmetry \cite{YamSug10}. 

The numerical wave functional obtained from the variational solution of Ref.~\cite{EppReiSch07}
seems to embody the correct
infrared physics as is revealed in the various applications considered to date: one finds a linearly
rising static quark potential \cite{EppReiSch07}, an infrared enhanced running coupling
constant with no Landau pole \cite{SchLedRei06}, a topological susceptibility in accord
with lattice data \cite{CamRei08}, a perimeter law for the 't Hooft loop \cite{ReiEpp07},
and, within an approximate Dyson--Schwinger equation, an area law for the spatial Wilson
loop \cite{PakRei09}. Furthermore, in Ref.\ \cite{Rei08} it was shown that the inverse
ghost form factor of Coulomb gauge Yang--Mills theory represents the dielectric function
of the Yang--Mills vacuum and the so-called horizon condition \cite{Zwa98} (of an infrared diverging
ghost form factor) implies that the Yang--Mills vacuum is a perfect color dielectricum,
i.e., a dual superconductor, which establishes the connection between the Gribov--Zwanziger
confinement scenario \cite{Gri78,Zwa98} and the monopole condensation picture \cite{Man76,tHo81}.
Finally, in Ref.\ \cite{Led+10} the functional renormalization group flow equation of
the Hamiltonian approach to Coulomb gauge Yang--Mills theory was studied, yielding
results for the gluon and ghost propagator similar to that of the variational approach
\cite{FeuRei04}. 

In the present paper, we generalize the variational approach to the Hamiltonian formulation
of Yang--Mills theory to non-Gaussian wave functionals. We will present a general method
to treat non-Gaussian wave functionals in quantum field theory. The method is based on
the observation that expectation values in the Hamiltonian formulation of $d = 3 + 1$
dimensional quantum field theory can be formally obtained from a generating functional
of $d = 3$ dimensional Euclidean quantum field theory with an action defined by the
logarithm of the vacuum wave functional. Expanding this action functional in powers
of the underlying field results in ``bare'' $n$-point kernels $\gamma_n$ as expansion
``coefficients''. We then exploit Dyson--Schwinger equation (DSE) techniques \cite{Dys49,Sch51,ItzZubBook}
to express the expectation value of the Hamiltonian $\vev{H}$ in terms of these
kernels $\gamma_n$, which are then determined by the variational principle, i.e., by minimizing
$\vev{H}$. This approach is then applied to the Hamiltonian formulation of
Yang--Mills theory in Coulomb gauge to include three- and four-gluon interaction kernels
in the exponent of the Yang--Mills vacuum wave functional.

By using such a non-Gaussian wave functional, the gluon loop is retained in the
expectation value of the Hamiltonian. Although the gluon loop is irrelevant for the
IR properties, it certainly influences the mid-momentum and UV regime of the gluon
propagator and thus of the running coupling, and also contributes to the anomalous
dimensions. As a first estimate of the effects of the non-Gaussian terms in the wave
functional, we will calculate the gluon-loop contribution to the gluon propagator as
well as the three- and four-gluon proper vertices using the ghost and gluon propagators
obtained from the Gaussian wave functional \cite{EppReiSch07} as input. A full
self-consistent inclusion of the three- and four-gluon vertices will be the
subject of future research.

It is clear from the very beginning that eventually we have to truncate the tower of
DSEs for the proper $n$-point vertex functions $\Gamma_n$ as well as the equations of
motion for the variational kernels $\gamma_n$ following from the variational principle.
For a systematic counting of the various diagrams we will assume a skeleton expansion.

Previous variational calculations (using Gaussian wave functionals) were restricted to
two (overlapping) loops in the energy $\vev{H}$, resulting in a one-loop gap equation
for the gluon propagator. Restriction to two overlapping loops in the energy results
in a bare (zero-loop) three-gluon kernel $\gamma_3$ and in a vanishing four-gluon
kernel, $\gamma_4=0$. To get a $\gamma_4\neq0$, one has to include up to three loops in
the energy. To keep the calculation sufficiently simple, we will keep only those three
(overlapping) loop terms in the energy containing three- or four-gluon kernels. This
will result in a bare (zero-loop) four-gluon and a one-loop three-gluon vertex.

The organization of the paper is as follows: in Sec.~\ref{sec:dse-gen} we present the
DSEs of the Hamiltonian approach  first for a general field theory and afterwards for
Yang--Mills theory in Coulomb gauge. The full static
(equal-time) propagators of the Hamiltonian approach are expressed in terms of proper vertex
functions in Sec.~\ref{sec:dec}. In Sec.~\ref{sec:dse} we specify our Yang--Mills vacuum
wave functional and derive the corresponding DSEs for the gluon and ghost proper $n$-point
functions. By means of these DSEs, the vacuum expectation value of the Hamiltonian is
expressed in Sec.~\ref{sec:energy} in terms of the variational kernels of the vacuum
wave functional. In Sec.~\ref{sec:var} these kernels are determined by minimizing the
energy density. Finally, in Sec.~\ref{sec:vertices} we calculate the three- and four-gluon
proper vertices using as input the ghost and gluon propagators from the
variational calculations with a Gaussian wave functional. A short summary and our conclusions
are given in Sec.~\ref{sec:summary}.


\section{\label{sec:dse-gen}Dyson--Schwinger equations of the Hamiltonian approach to
Yang--Mills theory in Coulomb gauge}

\subsection{General DSE formalism of the Hamiltonian approach to quantum field theory}

Consider a quantum field theory comprised of a collection of fields $\phi = ( \phi_1, \phi_2,
\dots )$ and let $\ket{\psi}$ be the exact vacuum state. All static (time-independent)
Green's functions, i.e., vacuum expectation values $\langle \phi \phi \dots \rangle$,
can be calculated from the generating functional
\be
\label{G1}
Z [j] = \bra{\psi}  \e^{\int j \cdot \phi} \ket{\psi} \: ,
\ee
where $j = ( j_1, j_2, \dots )$ stands for the collection of sources corresponding to
the fields and we use the abbreviation $j \cdot \phi = j_1 \phi_1 + j_2 \phi_2 + \cdots$.
In the ``coordinate'' representation of the vacuum state $\braket{\phi}{\psi} = \psi[\phi]$,
the scalar product in (\ref{G1}) is defined by the functional integral over time-independent
fields $\phi(\vx)$
\be
\label{G2}
Z [j] = \int \calD \phi \: \abs{\psi[\phi]}^2 \: \e^{\int j \cdot \phi} .
\ee
Furthermore, the integral in the exponent is over spatial coordinates $\vx$ of the
static fields $\phi(\vx)$.
Expressing the vacuum wave functional in the form%
\footnote{As long as we ignore the $\theta$--vacuum of Yang--Mills theory, the vacuum
wave functional can be chosen to be real, which we will assume in the present paper.}
\be
\label{G3}
\psi[\phi] = \exp \left( - \frac{1}{2} \, S[\phi] \right) ,
\ee
the generating functional of the Hamiltonian approach to quantum field theory becomes
\be
\label{G4}
Z [j] =  \int \calD \phi \: \e^{- S[\phi] + \int j \cdot \phi} ,
\ee
which is a standard generating functional of the $d=3$-dimensional Euclidean quantum field
theory defined by an ``action'' $S[\phi]$. Here, this action is defined by the vacuum
wave functional $\psi[\phi]$ and will, in general, be non-local and non-linear. We
therefore perform a Taylor expansion of the action functional $S[\phi]$ in powers of
the time-independent fields $\phi(\vx)$. The constant part $S[0]$ is fixed by the
normalization of the wave functional and the linear part can be absorbed into the
external source. It is then sufficient to consider expansions of $S[\phi]$ starting
at second order
\be
\label{G5}
S[\phi] = \frac{1}{2} \int \gamma_2 \: \phi^2 + \frac{1}{3!} \int \gamma_3 \: \phi^3 + \cdots \; .
\ee
Restricting the expansion to second order yields a Gaussian wave functional (\ref{G3})
for which the functional integral in Eq.~(\ref{G2}) can be explicitly carried out.
This corresponds to the so-called mean-field approximation, where all higher order
Green's functions of the field $\phi(\vx)$ are given in terms of the propagator $\vev{\phi \phi}$. 

In many cases the mean-field approximation is, however, not sufficient. Going beyond
the mean-field approximation, the functional integral in (\ref{G2}) can no longer be
explicitly performed. However, we can calculate the desired Green functions by
exploiting Dyson--Schwinger equation techniques. Starting from the identity
\be
\label{G6}
\int \calD \phi \: \frac{\delta}{\delta \phi} \left( \e^{- S[\phi] + \int j \phi} \right) = 0 
\ee
we can derive, in the standard fashion, a set of Dyson--Schwinger equations (DSEs) for
the Green functions $\vev{\phi \phi \cdots}$. This infinite tower of equations has to
be truncated to get a closed system of equations, and further simplifying assumptions
on the form of the interaction kernels $\gamma_n$ entering the ansatz for the vacuum
wave functional, see Eqs.~(\ref{G3}) and (\ref{G5}), will be required. Nevertheless,
this approach allows us to go beyond Gaussian wave functionals and calculate the static
Green functions $\vev{\phi \phi \cdots}$ in terms of the kernels $\gamma_n$. By
means of these static Green functions, the vacuum expectation value of the Hamiltonian
$\bra{\psi} H \ket{\psi}$ is expressed in terms of the kernels $\gamma_n$, which are
then found by minimizing the energy density. 

In the Hamiltonian approach to quantum field theory one is not primarily interested in
the generating functional \eqref{G2} itself but in expectation values of observables,
in particular of the Hamiltonian. For this purpose it turns out to be more convenient
to generalize Eq.~\eqref{G6} to
\be\label{G6a}
\int \calD\phi \: \frac{\delta}{\delta \phi} \left( \e^{-S[\phi]} \: K[\phi] \right)
= \int \calD\phi \: \frac{\delta}{\delta \phi} \left( \psi^*[\phi] \, K[\phi] \, \psi[\phi] \right) = 0 \, ,
\ee
where $K[\phi]$ is an arbitrary functional of the underlying field $\phi$.


\subsection{\label{kap3-1}Derivation of the DSEs for the Hamiltonian approach to
Yang--Mills theory in Coulomb gauge}

Below, we apply the general Dyson--Schwinger approach to the Hamiltonian formulation of
quantum field theory outlined above to Yang--Mills theory in Coulomb gauge (which
also assumes Weyl gauge $A_0^a=0$). Implementing
the Coulomb gauge by the Faddeev--Popov method, the expectation value of a functional
$K[A]$ of the (spatial components of the) gauge field $A$ is given by
\be\label{vev1}
\vev{K[A]} = \int_\Omega \calD A \: \calJ[A] \: \abs{\psi[A]}^2 \: K[A] \, .
\ee
Here, $\psi[A] = \braket{A}{\psi}$ denotes the Yang--Mills vacuum wave functional
restricted to transverse fields, $\partial_i A^a_i = 0$, and
$\calJ[A] = \Det (G^{- 1}_A)$ is the Faddeev--Popov determinant with
\be
\label{xyz}
G^{-1}_A = ( -\delta^{ab} \, {\partial^2_{\vx}} - g \, \hat{A}^{ab}_i(\vx) {\partial_i^{\vx}} ) \delta(\vx-\vy) \, 
\ee
being the Faddeev--Popov operator. Since we work only with spatial vectors, we will use
only Lorentz subscripts. Furthermore, $g$ is the coupling constant, $\hat{A}^{ab}=f^{acb} A^c$
is the gauge field in the adjoint representation of the colour group, and $f^{acb}$ are
the structure constants of the $\mathfrak{su}(N_c)$ algebra. The functional integration
in Eq.~(\ref{vev1}) runs over transverse field configurations and is restricted to the
first Gribov region $\Omega$ or, more precisely, to the fundamental modular region
\cite{Zwa94}. Moreover, we assume that the wave functional $\psi[A]$ is properly
normalized, $\braket{\psi}{\psi} \equiv \vev{1} = 1$. Writing the vacuum wave
functional as in Eq.~(\ref{G3})
\be\label{vac1}
\abs{\psi[A]}^2 =\mathrel{\mathop:} \e^{-S[A]}
\ee
and choosing
\be
\label{F2-XX}
K [A] = \e^{\int j \cdot A} \, ,
\ee
Eq.~(\ref{vev1}) becomes the generating functional of the static Green functions of
the (transverse) gauge field $A$. In the following, it will be convenient not to fix
$K[A]$ to the form (\ref{F2-XX}) but rather to let $K [A]$ be an arbitrary functional
of the gauge field. Furthermore, to simplify the bookkeeping we will use the compact
notation
\be
\label{183-1}
A^{a_1}_{k_1} (\vx_1)  =  A (1) \, , \quad
A \cdot B  =  A(1) \, B(1) = \int \d[d]x \: A_i^a(\vx) \, B_i^a(\vx) \, ,
\ee
such that a repeated label means summation over the discrete colour and Lorentz indices
along with integration over the spatial coordinates. 

Consider now the following identity
\be
\label{191-1}
0 = \int_\Omega \calD A \: \frac{\delta}{\delta A(1)} \left\{ \calJ[A] \: \e^{-S[A]} \: K[A] \right\} ,
\ee
which holds due to the fact that the Faddeev--Popov determinant $\calJ[A]$ vanishes on
the Gribov horizon $\partial\Omega$, tacitly assuming that the considered functional
$K[A]$ does not spoil the vanishing of $\calJ[A] K[A]$ on $\partial\Omega$.
Eq.~\eqref{191-1} with $K[A]$ given by Eq.~\eqref{F2-XX} becomes the ordinary DSE of
the usual (Lagrangian based) functional integral formulation of Yang--Mills theory in
Coulomb gauge \cite{WatRei07b} when the functional integration is extended over time-dependent gauge
fields $A_\mu(\vx,t)$ and $S[A]$ is chosen as the usual classical action of Yang--Mills
theory.

Working out the functional derivative in Eq.~\eqref{191-1} yields the following identity
\be\label{dserough}
\vev{\left[ \frac{\delta\ln\calJ}{\delta A(1)} - \frac{\delta S[A]}{\delta A(1)} \right] K[A]} +
\vev{\frac{\delta K[A]}{\delta A(1)}} = 0 \, .
\ee
The derivative of $\ln \calJ$ can be written as
\be\label{2-6x}
\frac{\delta\ln\calJ}{\delta A(1)} = \frac{\delta}{\delta A (1)} \Tr \ln G^{- 1}_A 
= \widetilde{\Gamma}_0(1;3,2) \, G_A(2,3) \, ,
\ee
where we have introduced the bare ghost-gluon vertex\footnote{%
The bare ghost-gluon vertex $\widetilde{\Gamma}_0$ defined by Eq.~(\ref{5-16}) differs
from the one of Ref.~\cite{FeuRei04} by an overall sign.} 
\be
\label{5-16}
\widetilde{\Gamma}_0 (1; 2, 3) = \frac{\delta G^{-1}_A(2, 3)}{\delta A(1)} \, .
\ee
With this result, Eq.~(\ref{dserough}) can be cast in the form
\be\label{dse}
\vev{\frac{\delta S[A]}{\delta A(1)} \: K[A]} = \vev{\frac{\delta K[A]}{\delta A(1)}}
+ \widetilde{\Gamma}_0(1;3,2) \vev{G_A(2,3) \: K[A]},
\ee
which is the basis of the gluon DSEs, exploited below in the evaluation of $\vev{H}$.

Introducing ghost fields in the usual way
\be\label{ghostdef}
\calJ[A] = \Det(G^{-1}_A) = \int\calD\bar{c} \: \calD c \: \e^{-\bar{c} G^{-1}_A c} \, ,
\ee
the expectation value \eqref{vev1} explicitly reads
\be\label{vev1g}
\vev{K[A]} = \int_\Omega \calD A \int \calD\bar{c} \: \calD c \: K[A] \:
\e^{-S[A]-\bar{c} G^{-1}_A c } \, ,
\ee
and Eq.~(\ref{dse}) can be written as
\be\label{dse-ghosts}
\vev{\frac{\delta S[A]}{\delta A(1)} \: K[A]} = \vev{\frac{\delta K[A]}{\delta A(1)}}
+ \widetilde{\Gamma}_0(1;3,2) \vev{c(2) \, \bar{c}(3) \, K[A]}.
\ee
The bare vertex $\widetilde{\Gamma}_0$ is the lowest-order perturbative contribution
\cite{CamReiWeb09} to the full ghost-gluon vertex $\widetilde{\Gamma}$ defined by
\be\label{ggvdef}
\vev{A(1) \, G_A(2,3)} = \vev{A(1) \, c(2) \, \bar{c}(3)}
= - D(1,1') \, G(2,2') \,  \widetilde{\Gamma}(1';2',3') \, G(3',3) \, ,
\ee
where
\be
\label{rrr}
D (1, 2) = \vev{ A(1) \, A(2)}
\ee
is the gluon propagator and
\be
\label{246-1}
G(1, 2) \mathrel{\mathop:}= \vev{ G_A(1,2) } = \vev { c(1) \, \bar{c}(2) } \, .
\ee
is the ghost propagator.

Eq.~\eqref{dse-ghosts} (or equivalently Eq.~\eqref{dse}) is the basic DSE of the Hamiltonian
formulation of Yang--Mills theory in Coulomb gauge, and we will refer to it as ``Hamiltonian
DSE''. Below we will exploit this equation to express the various
static (equal-time) correlators occurring in the vacuum expectation value of the Hamilton
operator by the variational kernels $\gamma_n$, Eq.~\eqref{G5}, of the wave functional
$\psi[A]$. This requires appropriate choices of the so far arbitrary functional $K[A]$.


\section{\label{sec:dec}Expressing static correlators through propagators and proper vertex functions}
\begin{figure}
\includegraphics{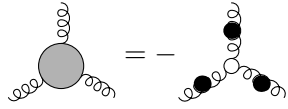}
\caption{\label{fig:3gvdec}Expression of the gluon three-point function, Eq.~\protect\eqref{tgv},
by means of three-point proper vertex function and propagators. Here and in the following,
fat shaded gray blobs represent full Green's functions, small filled dots connected Green's functions,
and small empty dots stand for proper vertex functions.}
\end{figure}

\begin{figure}
\includegraphics{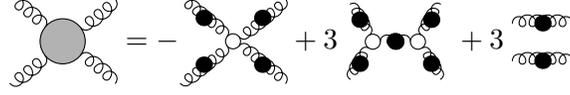}
\caption{\label{fig:4gvdec}Expression of the full gluon four-point function, Eq.~\protect\eqref{fgv}.
The prefactors indicate the number of possible permutations.}
\end{figure}

\begin{figure}
\includegraphics{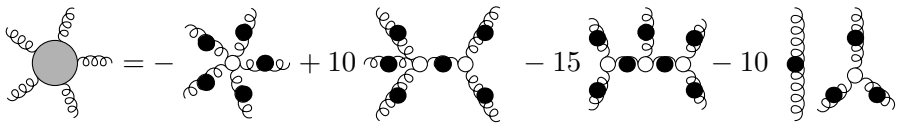}
\caption{\label{fig:5gvdec}Expression for the full gluon five-point Green function, Eq.~\protect\eqref{fivegv}.}
\end{figure}

Choosing the functional $K[A]$ in Eq.~\eqref{dse-ghosts} as
\be
\label{6-21}
K[A] = \exp \left\{ j \cdot A + \bar{c} \cdot \eta + \bar{\eta} \cdot c \right\} ,
\ee
where $j$ and $\bar{\eta}$, $\eta$ are the gluon and ghost sources, we obtain the
generating functional of the full static (equal-time) Green functions
\be
\label{6-22}
Z[j,\eta,\bar\eta] = \vev{\exp\{j A + \bar{c}\eta + \bar\eta c\}} =\mathrel{\mathop:} \e^{W[j,\eta,\bar\eta]} \, ,
\ee
where $W [j, \bar{\eta}, \eta]$ is the generating functional of the connected Green
functions
\be\label{conngf}
\begin{aligned}
\frac{\delta W}{\delta j(1)}\biggr|_{j=\bar{\eta}=\eta=0} &= \vev{A(1)} = 0 \, , &
\frac{\delta^2 W}{\delta j(1) \delta j(2)} \biggr|_{j=\bar{\eta}=\eta=0} &= \vev{A(1) A(2)} \, , \\
\frac{\delta^3 W}{\delta j(1) \delta \bar{\eta}(2) \delta\eta(3)} \biggr|_{j=\bar{\eta}=\eta=0} &= - \vev{A(1) c(2) \bar{c}(3)} \, ,
& &\text{etc.}
\end{aligned}
\ee
Introducing the classical fields as\footnote{%
With a slight abuse of notation, we employ the same symbol for both the classical
fields and the quantum fields which are integrated over. No confusion should arise,
since they never appear together. Furthermore, derivatives with respect to Grassmann
fields are always left derivatives.}
\be\label{cfields}
A = \frac{\delta W}{\delta j} \, , \quad
\bar{c} = - \frac{\delta W}{\delta \eta} \, , \quad
c = \frac{\delta W}{\delta \bar{\eta}} \, ,
\ee
we can define the effective action $\Gamma [A,\bar{c},c]$ through the Legendre transform
\be\label{legtrafo}
\Gamma[A,\bar{c},c] + W [j,\eta,\bar{\eta}] =
j \cdot A + \bar{c} \cdot \eta + \bar{\eta} \cdot c \, ,
\ee
where the sources have to be expressed by Eqs.~(\ref{cfields}) in terms of the classical
fields $A$, $\bar{c}$, $c$. From the effective action Eq.~\eqref{legtrafo}, the sources
are obtained as
\be\label{sources}
j = \frac{\delta\Gamma}{\delta A} \, , \quad
\eta =  \frac{\delta\Gamma }{\delta\bar{c}} \, , \quad
\bar{\eta} = - \frac{\delta\Gamma}{\delta c} \, .
\ee
Using Eqs.~\eqref{cfields}, differentiation with respect to the gluonic source can be
expressed as
\be\label{dersource}
\begin{split}
\frac{\delta}{\delta j(1)} &=
\frac{\delta A(2)}{\delta j(1)} \frac{\delta}{\delta A(2)} +
\frac{\delta c(2)}{\delta j(1)} \frac{\delta}{\delta c(2)} +
\frac{\delta \bar{c}(2)}{\delta j(1)} \frac{\delta}{\delta \bar{c}(2)} \\
&= \frac{\delta^2 W}{\delta j(1) \delta j(2)} \frac{\delta}{\delta A(2)} +
\frac{\delta^2 W}{\delta j(1) \delta\bar\eta(2)} \frac{\delta}{\delta c(2)} -
\frac{\delta^2 W}{\delta j(1) \delta\eta(2)} \frac{\delta}{\delta \bar{c}(2)} \, .
\end{split}
\ee
Similar expressions can be written for the derivatives with respect to the ghost sources.
Differentiating Eqs.~\eqref{sources} and using Eqs.~\eqref{cfields}, \eqref{dersource}
we can link, in the usual way, the connected Green functions (derivatives of $W$) with
the proper vertex functions (derivatives of $\Gamma$). As an example, we explicitly
show how to express the full ghost-gluon vertex $\widetilde{\Gamma}(1;2,3)$, defined in
Eq.~\eqref{ggvdef}, by derivatives of the effective action. We start from the
identity
\be\label{der-ggv-1}
\delta(1,2) =
\frac{\delta}{\delta\eta(1)} \frac{\delta\Gamma}{\delta\bar{c}(2)} = 
\frac{\delta^2 W}{\delta\eta(1) \delta j(2')} \frac{\delta^2\Gamma}{\delta A(2')\delta\bar{c}(2)} +
\frac{\delta^2 W}{\delta\eta(1) \delta\bar\eta(2')} \frac{\delta^2\Gamma}{\delta c(2')\delta\bar{c}(2)} -
\frac{\delta^2 W}{\delta\eta(1) \delta\eta(2')} \frac{\delta^2\Gamma}{\delta \bar{c}(2')\delta\bar{c}(2)} \, ,
\ee
which can be derived along the line of Eq.~\eqref{dersource}. Differentiating
Eq.~\eqref{der-ggv-1} with respect to a gluonic source $j(3)$ and using Eq.~\eqref{dersource}
yields
\be\label{der-ggv-2}
0 = \frac{\delta^3 W}{\delta j(3) \delta\eta(1) \delta\bar\eta(2')}
\frac{\delta^2\Gamma}{\delta c(2') \delta\bar{c}(2)}
+
\frac{\delta^2 W}{\delta\eta(1)\delta\bar\eta(2')}
\frac{\delta^2 W}{\delta j(3) \delta j(3')}
\frac{\delta^3 \Gamma}{\delta A(3') \delta c(2') \delta\bar{c}(2)}
+ \ldots
\ee
where the omitted terms vanish when the sources are set to zero. By means of Eq.~\eqref{conngf}
and of
\be
\frac{\delta^2\Gamma[A,\bar{c},c]}{\delta c(2) \delta\bar{c}(1)}\biggr|_{A=\bar{c}=c=0} = G^{-1}(1,2) \, ,
\ee
the last relation can be expressed as
\be\label{der-ggv-3}
0 = - G^{-1}(2,2') \vev{A(3) \bar{c}(1) c(2')} + G(2',1) D(3,3') 
\frac{\delta^3\Gamma[A,\bar{c},c]}{\delta c(2') \, \delta\bar{c}(2) \, \delta A(3')}\biggr|_{A=\bar{c}=c=0} \, .
\ee
Comparison with Eq.~\eqref{ggvdef} shows
\be\label{263-1}
\widetilde{\Gamma}(1;2,3) =
\frac{\delta^3\Gamma[A,\bar{c},c]}
     {\delta c(3) \, \delta\bar{c}(2) \, \delta A(1)}
\biggr|_{A=\bar{c}=c=0} \, .
\ee
Similarly, defining the $n$-gluon proper vertex function by
\be
\label{269-1}
\Gamma_n \equiv \Gamma (1, 2, \dots, n) =
\frac{\delta^n \Gamma[A,\bar{c},c]}
     {\delta{A}(1) \, \delta{A}(2) \cdots \delta{A}(n)}
\biggr|_{A=\bar{c}=c=0} \, ,
\ee
the full gluon $n = 2,3,4,5$-point functions defined by Eq.~\eqref{vev1} with $K[A]=AA\ldots$ can be
expressed through the proper vertex functions as
\be\label{274-1}
D (1, 2) \equiv \vev{ A(1) \, A(2) } = \Gamma(1, 2)^{-1} \, ,
\ee
\be\label{tgv}
\vev{A(1) \, A(2) \, A(3)} = - \Gamma(1',2',3') \, D(1',1) \, D(2',2) \, D(3',3) \, ,
\ee
\be\label{fgv}
\begin{split}
& \langle A(1) \, A(2) \, A(3) \, A(4) \rangle = D(1,2) D(3,4) + D(1,3) D(2,4) + D(1,4) D(2,3) \\
& + D(1',1) \, D(2',2) \, D(3',3) \, D(4',4) \biggl\{
- \Gamma(1',2',3',4' )\\
& + D(5,5') \Bigl[ 
\Gamma(1',2',5) \Gamma(5',3',4') + \Gamma(1',3',5) \Gamma(5',2',4') + \Gamma(1',4',5) \Gamma(5',2',3') 
\Bigr]\biggr\} ,
\end{split}
\ee
\be\label{fivegv}
\begin{split}
\langle A(1) \, & A(2) \, A(3) \, A(4) \, A(5) \rangle = 
-\Gamma(1',\ldots,5') \, D(1,1') \ldots D(5,5') + \\
&+ \bigl[ \vev{A(1) A(2) A(6)}  \Gamma(6,3',4',5') D(3,3') D(4,4') D(5,5') + \text{9 combinations} \bigr] \\
&- \bigl[ D(1,1') \vev{A(1')A(6)A(7)} \Gamma(6,6') \Gamma(7,7') \vev{A(6')A(2)A(3)} \vev{A(7')A(4)A(5)} \\
{}& \qquad\qquad + \text{14 combinations} \bigr] \\
&- \bigl[ D(1,2) \vev{A(3) \, A(4) \, A(5)} + \text{9 combinations} \bigr] \, ,
\end{split}
\ee
The proper $n$-point gluonic functions $\Gamma(1, 2, \dots, n)$ \eqref{269-1} are by definition
invariant with respect to a permutation of external legs, i.e., of the entries $1, 2, \dots, n$.
Eqs.~\eqref{tgv}--\eqref{fivegv} are represented in diagrammatic form in
Figs.~\ref{fig:3gvdec}--\ref{fig:5gvdec}.%
The prefactors in Figs.~\ref{fig:4gvdec}, \ref{fig:5gvdec} indicate the number of the
possible combinations. Consider, e.g., the second diagram on the right-hand side of
Fig.~\ref{fig:5gvdec}: out of the five external legs one can form $\binom{5}{2} = 10$
pairs of external legs attached to the right vertex. The remaining three external legs
have to be attached then to the left vertex and thus do not add more possible combinations.
In the third diagram there are five possibilities to select the external leg attached
to the internal lines. From the remaining 4 external legs there are $\binom{4}{2}=6$
possibilities to choose the two external legs at the right external vertex. This fixes
also the external legs at the left vertex. The symmetry of the diagram with respect to
the interchange of the two external vertices introduces an extra factor $\frac{1}{2}$.
Therefore this diagram occurs with the multiplicity $5 \cdot 6 \cdot \frac{1}{2} = 15$.

\begin{figure}
\includegraphics{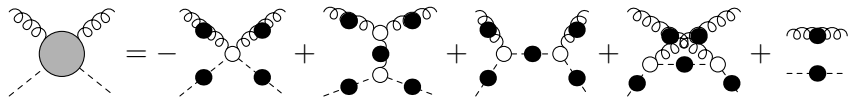}
\caption{\label{fig:2g2gvdec}Vacuum expectation value of two gauge fields and a ghost
and an anti-ghost field after Eq.~\protect\eqref{2g2gv}.}
\end{figure}

In a similar fashion one finds for the expectation value of two gauge fields and a
ghost and anti-ghost field
\be\label{2g2gv}
\begin{split}
\langle A(1) \, A(2) \, &c(3) \, \bar{c}(4) \rangle = 
\vev{A(1) \, A(2) \, G_A(3,4)} = 
D(1,2) \, G(3,4) \\
& + D(1',1) \, D(2',2) \, G(3,3') \, G(4',4) \biggl\{
- \widetilde\Gamma(1',2';3',4')\\
& + \Gamma(1',2',5) D(5,5') \widetilde{\Gamma}(5';3',4') \\
&+  \widetilde{\Gamma}(1';3',5) G(5,5') \widetilde{\Gamma}(2';5',4') +
\widetilde{\Gamma}(2';3',5) G(5,5') \widetilde{\Gamma}(1';5',4')
\Bigr]\biggr\} \, ,
\end{split}
\ee
where the two-gluon-two-ghost vertex is defined by
\be\label{2g2gv-1}
\widetilde\Gamma(1,2;3,4) =
\frac{\delta^4 \Gamma[A,\bar{c},c]}
     {\delta c(4) \, \delta\bar{c}(3) \, \delta A(2) \, \delta A(1)}
\biggr|_{A=\bar{c}=c=0} \, .
\ee
The diagrammatic representation of Eq.~\eqref{2g2gv} is shown in Fig.~\ref{fig:2g2gvdec}.

\begin{figure}
\includegraphics{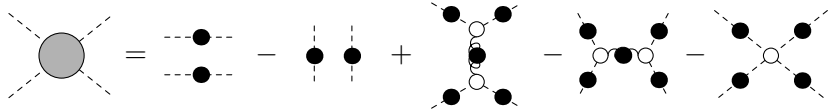}
\caption{\label{fig:4ghdec}Vacuum expectation value of two ghost and two anti-ghost
fields, Eq.~\protect\eqref{cff4}.}
\end{figure}

The last four-point function we need for the evaluation of $\vev{H}$ is the ghost four-point
function
\be\label{cff3}
\vev{c(1) \, \bar{c}(2) \, c(3) \, \bar{c}(4) } =
\vev{ \bigl[ G_A(1,2) \, G_A(3,4) - G_A(1,4) \, G_A(3,2) \bigr] } \, ,
\ee
which can be expressed in terms of propagators and proper vertices in the standard way,
yielding
\be\label{cff4}
\begin{split}
\vev{ c(1) \, \bar{c}(2) \, c(3) \, \bar{c}(4) } ={}& G(1,2) \, G(3,4) - G(1,4) \, G(3,2) + \\
& + G(1,1') \, G(3,3') \, \widetilde{\Gamma}(5;1',2') D(5,5') \widetilde{\Gamma}(5';3',4') \times \\
& \times[ G(2',2) \, G(4',4) - G(2',4) \, G(4',2) ] + \\
& - \widetilde{\Gamma}(1',3',2',4') \, G(1,1') \, G(2',2) \, G(3,3') \, G(4',4) \, ,
\end{split}
\ee
where the four-ghost vertex is defined by
\be\label{cff5a}
\widetilde\Gamma(1,3,2,4) =
\frac{\delta^4 \Gamma[A,\bar{c},c]}
     {\delta c(4) \, \delta c(2) \, \delta\bar{c}(3) \, \delta\bar{c}(1)}
\biggr|_{A=\bar{c}=c=0} \, .
\ee
Eq.~(\ref{cff4}) is represented diagrammatically in Fig.~\ref{fig:4ghdec}


\section{\label{sec:dse}The vacuum wave functional and corresponding DSEs}

So far, all manipulations have been exact. In Sec.~\ref{kap3-1} we have presented the
Hamiltonian DSEs for arbitrary wave functionals. To proceed further, we have to make
an ansatz for the form of the vacuum wave functional
$\psi[A]$, which by Eq.~\eqref{vac1} defines the ``action'' functional $S[A]$.

In perturbation theory, the vacuum wave functional in the form \eqref{G3}, \eqref{G5}
has been determined up to order $\calO(g^2)$ by a solution of the Schr\"odinger equation,
and the resulting expressions for the kernels $\gamma_2$, $\gamma_3$, and $\gamma_4$
are given in Ref.~\cite{Cam+09}. In the present non-perturbative approach, we will assume
a wave functional of the form \eqref{G3}, \eqref{G5} with an ``action'' functional to be given by
\be\label{ansatz}
\begin{split}
S[A] ={} &
\omega(1,2) \, A(1) \, A(2) + \frac{1}{3!} \: \gamma(1,2,3) \, A(1) \, A(2) \, A(3) +\\
&+ \frac{1}{4!} \: \gamma(1,2,3,4) \, A(1) \, A(2) \, A(3) \, A(4) .
\end{split}
\ee
For historical reason, we have denoted $\frac12 \gamma(1,2)$ by $\omega(1,2)$.
Consistent with our convention on the proper vertices Eq.~\eqref{269-1}, we will frequently
use the shorthand $\gamma_n\equiv\gamma(1,\ldots,n)$.
The functions $\omega\equiv\frac12\gamma_2$, $\gamma_3$, and $\gamma_4$ are variational kernels which will be
determined by minimization of the vacuum energy density. As discussed before,
Eq.~(\ref{ansatz}) can be considered as arising in leading orders of a systematic Taylor
expansion of the ``action'' functional.

Let us also stress that the ghost fields do not enter the Yang--Mills vacuum wave
functional $\psi[A]$. The ghost fields are auxiliary fields to represent the Faddeev--Popov
determinant in local action form. By the very definition of the ghost fields, Eq.~\eqref{ghostdef},
the ghost-gluon vertex in the ``action'' (i.e., the exponent of Eq.~\eqref{vev1g}) has
to be the bare vertex and, in principle, there is absolutely no need to include ghost
or ghost-gluon kernels as variational kernels in the wave functional. However, due to
approximations to be introduced, for practical purposes, one might also include ghost
or ghost-gluon vertices as variational kernels in the wave functional to improve the
latter. If the exact gluon wave functional $\psi[A]$ were used, the variational principle
would determine the ghost-gluon kernel as the bare one and higher ghost kernels to
vanish. Therefore we will not include additional ghost vertices into the variational
ansatz for the vacuum wave functional.

By construction, the variational kernels $\gamma_n$ (which are purely gluonic)
are totally symmetric with respect to permutations
of the overall indices. Furthermore, the wave functional defined by Eqs.~\eqref{vac1},
\eqref{ansatz} is normalizable even when the restriction of the functional integration
to the first Gribov region is ignored, provided the kernel $\gamma_4$ is positive definite,
which we will assume for the moment and which later on will be confirmed by our calculations.
With the action functional Eq.~\eqref{ansatz}, the Hamiltonian DSE~\eqref{dse} becomes
\be\label{dseansatz}
\begin{split}
&2 \omega(1,2) \vev{A(2) K[A]} + \frac12 \, \gamma(1,2,3) \vev{A(2) \, A(3) \, K[A]} + \\
&+ \frac{1}{3!} \, \gamma(1,2,3,4) \vev{A(2) \, A(3) \, A(4) \, K[A]}
= \vev{\frac{\delta K[A]}{\delta A(1)}} + \widetilde{\Gamma}_0(1;3,2) \vev{G_A(2,3) \: K[A]} \, .
\end{split}
\ee
Except for the non-locality of the variational kernels $\omega$, $\gamma_3$, and $\gamma_4$,
the functional Eq.~\eqref{ansatz} has the same structure as the ordinary Yang--Mills
action. Therefore, the DSEs resulting from Eq.~\eqref{dseansatz} will have the same
structure as the DSEs of ordinary $d = 3$ Yang--Mills theory in Landau gauge, however
with bare vertices replaced by the non-local variational kernels $\omega$, $\gamma_3$,
and $\gamma_4$. Eq.~\eqref{dseansatz} is our fundamental DSE for the Hamiltonian approach
to Yang--Mills theory in Coulomb gauge.


\subsection{\label{subsec:gluondses}DSEs of gluonic vertex functions}

\begin{figure}
\includegraphics{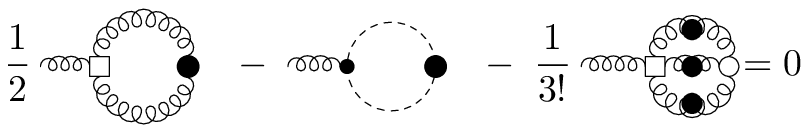}
\caption{\label{fig:tadpoleterms}Diagrammatic representation of Eq.~\protect\eqref{dse0}.
The empty square boxes denote the variational kernels $\gamma_{n}$.}
\end{figure}

The first DSE is obtained by setting $K[A]=1$ in Eq.~\eqref{dseansatz}. Using $\vev{A}=0$
and the expression Eq.~\eqref{tgv} for the three-point function yields the identity
\be\label{dse0}
\begin{split}
0 = {} & \frac12 \, \gamma(1,2,3) D(2,3) - \widetilde{\Gamma}_0(1;3,2) G(2,3) \\
& - \frac{1}{3!} \, \gamma(1,2,3,4) \, \Gamma(2',3',4') \, D(2,2') \, D(3,3') \, D(4,4') \, ,
\end{split}
\ee
which is diagrammatically illustrated in Fig.~\ref{fig:tadpoleterms}.
\begin{figure}
\includegraphics{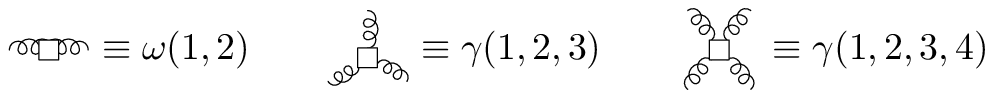}
\caption{\label{fig:kernels}Variational kernels occurring in the exponent of the wave functional.}
\end{figure}
Eq.~\eqref{dse0} is not really a dynamical equation but rather a constraint,
which can be used to simplify tadpole terms in the evaluation of higher-order DSEs. It
is also easy to see that in lowest order perturbation theory each term in Eq.~\eqref{dse0}
vanishes separately.

\begin{figure}
\includegraphics{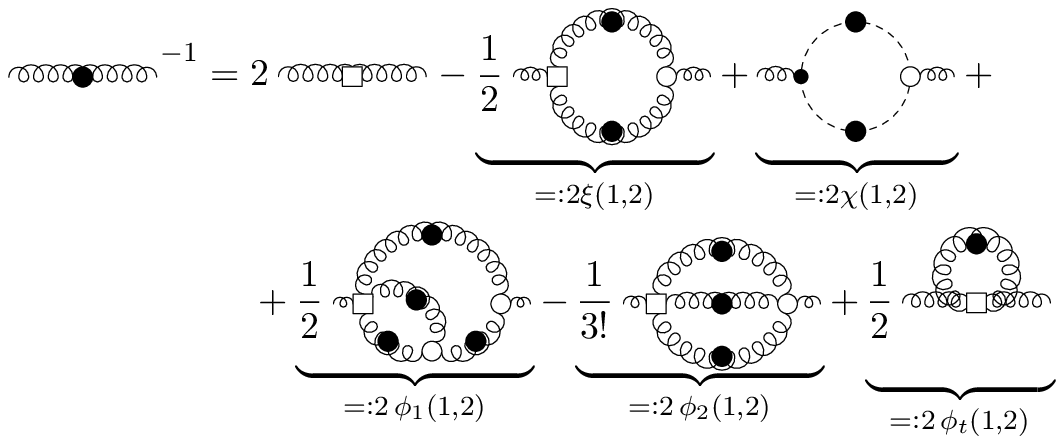}
\caption{\label{fig:gluonDSE}Diagrammatic representation of the DSE for the gluon
propagator, Eq.~\protect\eqref{dse10}.}
\end{figure}

The DSE for the gluon propagator follows from (\ref{dseansatz}) by putting $K[A]=A$,
yielding
\be\label{dse1}
\begin{split}
&2 \omega(1,3) \vev{A(3) \, A(2)} + \frac12 \, \gamma(1,3,4) \vev{A(3) \, A(4) \, A(2)} + \\
&+ \frac{1}{3!} \, \gamma(1,3,4,5) \vev{A(3) \, A(4) \, A(5) \, A(2)}
= t(1,2) + \widetilde{\Gamma}_0(1;4,3) \vev{G_A(3,4) \: A(2)} \, ,
\end{split}
\ee
where we have introduced the abbreviation
\be
t(1,2) \equiv \delta^{a_1a_2} \: t_{k_1 k_2}(\vx_1) \: \delta(\vx_1-\vx_2)
\ee
and $t_{k_1 k_2}(\vx_1) = \delta_{k_1 k_2}-\partial_{k_1}^{x_1} \partial_{k_2}^{x_1}/\partial^2_{x_1}$
is the transverse projector. By means of Eqs.~\eqref{tgv}, \eqref{fgv}, the three- and
four-point functions in Eq.~\eqref{dse1} can be expressed through the proper vertex
functions $\Gamma_n$. By multiplying Eq.~\eqref{dse1} by the inverse gluon propagator
Eq.~\eqref{274-1} and defining
\be\label{4-20}
D(1,2)^{-1} = \Gamma(1,2) =\mathrel{\mathop:} 2 \, \Omega(1,2) \, ,
\ee
Eq.~\eqref{dse1} can be cast in the form
\be\label{dse10}
\Omega(1,2) = \omega(1,2) - \xi(1,2) + \chi(1,2) + \phi_1(1,2) - \phi_2(1,2) + \phi_t(1,2) ,
\ee
where we have introduced the following loop terms
\begin{subequations}\label{dse10a}
\begin{align}
\xi(1,2) &= \tfrac{1}{4} \, \gamma(1,3,4) \, D(3,3') \, D(4,4') \, \Gamma(3',4',2) \label{dse10aa} , \\
\chi(1,2) &= \tfrac{1}{2} \, \widetilde{\Gamma}_0(1;3,4) \, G(3',3) \, G(4,4') \, \widetilde{\Gamma}(2;4',3') \label{dse10ab} , \\
\phi_1(1,2) &= \tfrac{1}{4} \, \gamma(1,3,4,5) \,
D(3,3') \, D(4,4') \, D(5,5') \, D(6,6') \,\Gamma(4',5',6) \, \Gamma(3',6',2) \label{dse10ac} , \\
\phi_2(1,2) &= \tfrac{1}{3! 2} \, \gamma(1,3,4,5) \, D(3,3') \, D(4,4') \, D(5,5') \, \Gamma(3',4',5',2) \label{dse10ad} , \\
\phi_t(1,2) &= \tfrac{1}{2} \, \gamma(1,2,3,4) \, D(3,4) \label{dse10ae} .
\end{align}
\end{subequations}
Eq.~\eqref{dse10} is represented diagrammatically in Fig.~\ref{fig:gluonDSE},
and is recognized as the usual DSE for the gluon propagator of Landau gauge Yang--Mills
theory \cite{MarPag78}, except for the replacement of the bare Yang--Mills vertices (defined by the
Yang--Mills Lagrangian) by the variational kernels $\gamma_n$ (defined by the ansatz
\eqref{ansatz} for the vacuum functional), which are represented by open square boxes,
see Fig.~\ref{fig:kernels}. Note that the gluon loop $\xi(1,2)$ Eq.~\eqref{dse10aa}
disappears when the three-gluon kernel $\gamma_3$ is absent from the exponential Eq.~\eqref{ansatz}
of the wave functional Eq.~\eqref{vac1}. For a Gaussian wave functional ($\gamma_3=\gamma_4=0$)
only the ghost loop $\chi(1,2)$ Eq.~\eqref{dse10ab} survives from the loop terms in the
DSE \eqref{dse10}, Fig.~\ref{fig:gluonDSE}.

\begin{figure}
\includegraphics{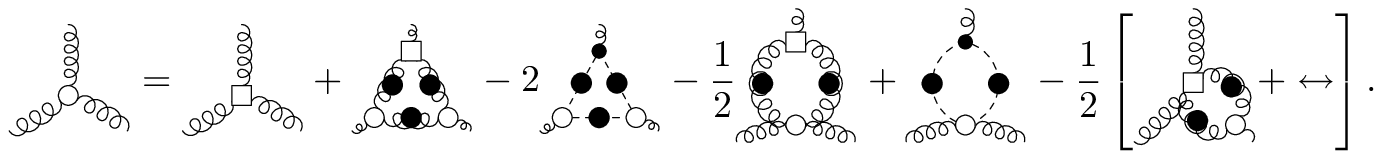}
\caption{\label{fig:3gvDSE}Diagrammatic representation of the DSE \protect\eqref{dse11}
for the three-gluon proper vertex function. The factor 2 in front of the ghost loop
accounts for the two diagrams differing in the direction of the ghost line.}
\end{figure}

Choosing $K[A] = A(2) A(3)$ in Eq.~\eqref{dseansatz} yields the DSE for the
three-gluon vertex,
\be\label{dse11a}
\begin{split}
&2 \omega(1,4) \vev{A(4) A(2) A(3) } + \frac12 \, \gamma(1,4,5) \vev{A(4) \, A(5) \, A(2) \, A(3) } + \\
&+ \frac{1}{3!} \, \gamma(1,4,5,6) \vev{A(4) \, A(5) \, A(6) \, A(2) \, A(3)}
= \widetilde{\Gamma}_0(1;5,4) \vev{G_A(4,5) \: A(2) \, A(3)} \, .
\end{split}
\ee
By means of Eqs.~(\ref{tgv})--(\ref{2g2gv}), the first two terms on the left-hand side
and the right-hand side of Eq.~(\ref{dse11a}) can be expressed in terms of proper vertex
functions $\Gamma_n$. The explicit evaluation of the five-point function is quite lengthy, and
we quote only the result. Restricting ourselves to terms involving up to one loop (which
is sufficient to obtain the energy $\vev{H}$ up to three overlapping loops, see the
introduction), and
chopping off the external propagators we eventually find from Eq.~\eqref{dse11a} the
DSE for the proper three-point vertex function $\Gamma(1,2,3)$
\be\label{dse11}
\begin{split}
\Gamma(1,2,3) ={}& \gamma(1,2,3) + \gamma(1,4,5) \, D(4,4') \, D(5,5') \, D(6,6') \, \Gamma(2,4',6) \, \Gamma(3,5',6') \\
&- \widetilde{\Gamma}_0(1;4,5) \, G(4',4) \, G(5,5') \, G(6',6) \bigl[
 \widetilde{\Gamma}(2;6,4') \, \widetilde{\Gamma}(3;5',6') + 2 \leftrightarrow 3 \bigr] \\
&-\frac12 \: \gamma(1,4,5) \, D(4,4') \, D(5,5') \, \Gamma(4',5',2,3) \\
&+ \widetilde{\Gamma}_0(1;4,5) \, G(4',4) \, G(5,5') \, \widetilde{\Gamma}(2,3;5',4') \\
&- \frac12 \bigl[ \gamma(1,2,4,5) \, D(4,4') \, D(5,5') \, \Gamma(4',5',3) + 2 \leftrightarrow 3 \bigr] ,
\end{split}
\ee
which is represented diagrammatically in Fig.~\ref{fig:3gvDSE}.

Analogously, one can derive the DSE for the four-gluon vertex. Restricting ourselves
again up to three loops in the energy $\vev{H}$, one finds just the ``tree-level''
expression
\be\label{dse12}
\Gamma(1,2,3,4) = \gamma(1,2,3,4) + \dots
\ee
Any loop contribution to $\Gamma_4$ generates at least four-loop terms in the energy,
which are beyond the scope of the present paper.
\begin{figure}
\includegraphics{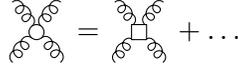}
\caption{\label{fig:4gvDSE}Leading order DSE for the four-gluon vertex.}
\end{figure}


\subsection{\label{subsec:ghdse}The DSEs for the ghost propagator and the ghost-gluon vertex}
\begin{figure}
\includegraphics{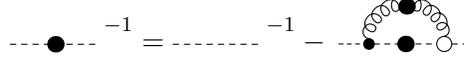}
\caption{\label{fig:ghostDSE}Diagrammatic representation of the ghost DSE \protect\eqref{ghdse2}.}
\end{figure}

Inverting the defining equation of the Faddeev--Popov operator \eqref{xyz} one finds
the following identity \cite{FeuRei04}
\be\label{ghdse1}
G_A(1,2) = G_0(1,2) - G_A(1,4) \, A(3) \, \widetilde{\Gamma}_0(3;4,5) \, G_0(5,2) \, ,
\ee
where $G_0(1,2) = [(-\partial^2)^{-1}](1,2)$ is the bare ghost propagator and
$\widetilde{\Gamma}_0$ is the bare ghost-gluon vertex defined in Eq.~\eqref{5-16}.
Taking the expectation value of Eq.~\eqref{ghdse1} and using Eq.~\eqref{ggvdef} yields
the usual DSE for the ghost propagator \cite{FeuRei04}
\be\label{ghdse2}
G(1,2)^{-1} = G_0(1,2)^{-1} -
\widetilde{\Gamma}(3;1,4) \, G(4,4') \, D(3,3') \, \widetilde{\Gamma}_0(3';4',2) \, ,
\ee
which in momentum space reads
\be\label{ghdsemom}
G^{-1}(\vp) = \vp^2 + \frac{\i \, g}{N_c^2-1} \int \frac{\d[d]q}{(2\pi)^d} \:
f^{abc} \widetilde{\Gamma}_i^{abc}(\vq;\vp-\vq,\vp) \: \frac{t_{ij}(\vp) \, p_j}{2 \Omega(\vq)} \: G(\vp-\vq) \, ,
\ee
and which is represented diagrammatically in Fig.~\ref{fig:ghostDSE}.

\begin{figure}
\includegraphics{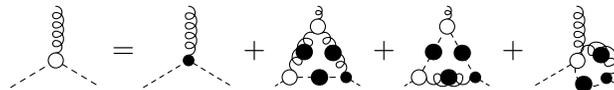}
\caption{\label{fig:ggvDSE1}Diagrammatic representation of the DSE \protect\eqref{ghdse3}
for the ghost-gluon vertex.}
\end{figure}

\begin{figure}
\includegraphics{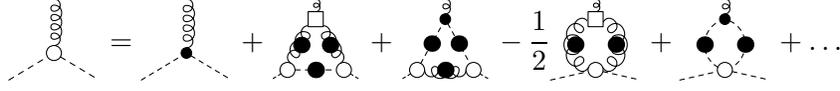}
\caption{\label{fig:ggvDSE2}Diagrammatic representation of the DSE \protect\eqref{ghdse5}
for the ghost-gluon vertex.}
\end{figure}

To derive the DSE for the ghost-gluon vertex there are two possibilities. The first one
is to multiply Eq.~\eqref{ghdse1} by the gauge field and to take the expectation
value of the resulting expression. This leads to
\be\label{ghdse3}
\vev{A(1) \, G_A(2,3)} = - \vev{A(1) \, A(4) \, G_A(2,5)} \,
\widetilde{\Gamma}_0(4;5,6) \, G_0(6,3) \, .
\ee
The remaining expectation value can be expressed in terms of proper vertices by means
of Eqs.~\eqref{ggvdef} and \eqref{2g2gv}. After chopping off the external propagators,
this results in the DSE for the ghost-gluon vertex shown in Fig.~\ref{fig:ggvDSE1}.
Eq.~\eqref{ghdse3} is exact, i.e., not truncated, but not very convenient for the
evaluation of the energy density. A more convenient form of the DSE for the ghost-gluon
vertex is obtained by putting $K[A]=G_A$ in our general Hamiltonian DSE \eqref{dseansatz},
thereby using the chain rule for the derivative of the ghost Green's function and using
Eqs.~\eqref{2g2gv}, \eqref{cff4} to express vacuum expectation values through propagators
and proper vertex functions. The resulting equation reads at one-loop level
\be\label{ghdse5}
\begin{split}
\widetilde{\Gamma}(1;2,3) ={}& \widetilde{\Gamma}_0(1;2,3)
+ \gamma(1,4,5) \, D(4,4') \, D(5,5') \, G(6,6') \, \widetilde{\Gamma}(4';2,6) \, \widetilde{\Gamma}(5';6',3) \\
&+ \widetilde{\Gamma}_0(1;4,5) \, G(4',4) \, G(5,5') \, D(6,6') \, \widetilde{\Gamma}(6,2,4') \, \widetilde{\Gamma}(6';5',3) \\
&-\frac12 \: \gamma(1,4,5) \, D(4,4') \, D(5,5') \, \Gamma(4',5';2,3) \\
&+ \widetilde{\Gamma}(1;4,5) \, G(4',4) \, G(5,5') \, \widetilde{\Gamma}(2,5',3,4') \, .
\end{split}
\ee
and is shown in Fig.~\ref{fig:ggvDSE2}.
The main difference between Eqs.~\eqref{ghdse3} and \eqref{ghdse5} is that while
Eq.~\eqref{ghdse3} is exact, in Eq.~\eqref{ghdse5} two-loops terms involving higher-order
vertices are neglected. Nevertheless, to our purpose, calculating the energy up to three
loops, Eq.~\eqref{ghdse5} is more convenient. We will use this equation in Sec.~\ref{sec:energy}
to simplify the expression for the kinetic and Coulomb energy.


\section{\label{sec:energy}Energy density of the Yang--Mills vacuum}

The DSEs of the Hamiltonian approach derived in Sec.~\ref{sec:dse} are not ``equations
of motion'' in the usual sense, but rather connect the various Green functions with the
kernels occurring in the ansatz for the wave functional, while these kernels themselves
are at this point not yet fixed. Here is where the variational principle comes into
play: we will now evaluate the expectation value of the Yang--Mills Hamiltonian for the
wave functional (\ref{vac1}) with the ansatz (\ref{ansatz}) and then minimize it with
respect to the variational kernels $\gamma_n$.

The Yang--Mills Hamiltonian in Coulomb gauge reads \cite{ChrLee80}
\be\label{609-1}
\begin{split}
H ={} & \int \d[d]x 
\left[ \frac{1}{2} \: \calJ^{-1}[A] \, \Pi^a_i (\vx) \, \calJ[A] \, \Pi^a_i (\vx) +
\frac{1}{4} \: F^a_{ij}(\vx) \, F^a_{ij}(\vx) \right] \\
& + \frac{g^2}{2} \int \d[d]x \, \d[d]y \: \calJ^{- 1}[A] \, \rho^a(\vx) \,
F^{ab}_A (\vx, \vy) \, \calJ[A] \, \rho^b(\vy) \, .
\end{split}
\ee
Here, $\Pi^a_k (\vx) = - \i \delta / \delta A^a_k (\vx)$ is the momentum operator, and
$F^a_{ij} = \partial_i A^a_j - \partial_j A^a_i + g f^{abc} A^b_i A^c_j$, is the
non-Abelian field strength tensor. The first two terms in Eq.~\eqref{609-1} are the
electric (kinetic) and magnetic parts of the ordinary Yang--Mills Hamiltonian restricted to the
curvilinear ``coordinate'' space of Coulomb gauge. The third (Coulomb) term arises
from the resolution of Gauss's law and describes the interaction of the non-Abelian
colour charge (of the fluctuating gauge field) with density
\be
\label{615-1}
\rho^a (\vx) = \hat{A}^{ab}_i(\vx) \, \Pi^b_i(\vx)
\ee
through the non-Abelian Coulomb interaction kernel
\be
\label{620-1}
F^{ab}_A(\vx,\vy) =
\bigl[ (- \hat{D} \partial)^{- 1} (- \partial^2) (- \hat{D} \partial)^{- 1} \bigr]_{\vx,\vy}^{a,b} \: .
\ee
The Yang--Mills Hamiltonian in Coulomb gauge Eq.~\eqref{609-1} is a positive definite operator.
Accordingly the energy $\bra{\psi} H \ket{\psi}$ is bounded from below (by zero) and
the variational principle is applicable.

For later use, we rewrite the magnetic term of the Hamiltonian in the symmetrized form 
\be\label{magham}
\frac14 \, F_{ij}^a \, F_{ij}^a =
-\frac12 A \partial^2 A + \frac{g}{3!} \, T_3 A^3 + \frac{g^2}{4!} \, T_4 A^4 \, ,
\ee
where the interaction kernels are given in momentum space by
\begin{subequations}\label{maghamt}
\be\label{t3}
T^{abc}_{ijk}(\vp,\vq,\vk) = \i \, f^{abc}
[ \delta_{ij} (p-q)_k + \delta_{jk} (q-k)_i + \delta_{ki} (k-p)_j ]
\ee
and
\be\label{t4}
\begin{split}
T^{abcd}_{ijkl} =
\bigl\{ {}&
f^{abe} f^{cde} (\delta_{ik} \, \delta_{jl} - \delta_{il} \, \delta_{jk}) \\
+{}& f^{ace} f^{bde} (\delta_{ij} \, \delta_{kl} - \delta_{jk} \, \delta_{il}) \\
+{}& f^{ade} f^{bce} (\delta_{ij} \, \delta_{kl} - \delta_{ik} \, \delta_{jl})
\bigr\} .
\end{split}
\ee
\end{subequations}
The bare four-gluon vertex, Eq.~(\ref{t4}), is independent of the momenta.


\subsection{\label{subsec:tec}Technicalities}
To evaluate the vacuum expectation values of the kinetic term and of the Coulomb
Hamiltonian, it is convenient to perform an integration by parts in the gauge field.
This leads to expressions of the form
\be\label{ke1}
\vev{\frac{\delta S[A]}{\delta A(1)} \: \frac{\delta S[A]}{\delta A(2)} \: f[A] },
\ee
where $f[A]$ is a functional of the gauge field which does not contain any momentum
operator. In principle, we could now explicitly write down the variations of the action
\eqref{ansatz} and evaluate the expectation value \eqref{ke1}. Then one would recognize
that some terms can be combined and simplified by using the DSEs \eqref{dse10},
\eqref{dse11}. Therefore, a more efficient way to evaluate the expectation value
\eqref{ke1} is to use the Hamiltonian DSEs from the very beginning. 

Putting $K[A]=\delta S/\delta A(2) f[A]$ in the general DSE (\ref{dse}) we obtain
\be\label{tec2}
\begin{split}
\vev{\frac{\delta S[A]}{\delta A(1)} \: \frac{\delta S[A]}{\delta A(2)} \: f[A]}
={} &
\vev{\frac{\delta^2 S[A]}{\delta A(1) \, \delta A(2)} \: f[A] } \\
& + \vev{\frac{\delta f[A]}{\delta A(1)} \: \frac{\delta S[A]}{\delta A(2)} }
+ \widetilde{\Gamma}_0(1;3,4) \vev{\frac{\delta S[A]}{\delta A(2)} \: G_A(4,3)} \, .
\end{split}
\ee
The last two terms on the r.h.s.\ of Eq.~(\ref{tec2}) can again be re-expressed through
the DSEs (\ref{dse}), thereby putting $K[A]=\delta f[A]/\delta A$ and $K[A]=G_A$, respectively,
and using the definition of the bare ghost-gluon vertex $\widetilde{\Gamma}_0$, Eq.~\eqref{5-16}.
This results finally in the relation
\be\label{ke3}
\begin{split}
\biggl<\frac{\delta S[A]}{\delta A(1)} \: & \frac{\delta S[A]}{\delta A(2)} \: f[A] \biggr>=
\vev{\frac{\delta^2 S[A]}{\delta A(1) \, \delta A(2)} \: f[A]}
+\vev{\frac{\delta^2 f[A]}{\delta A(1) \, \delta A(2)}} \\
& + \widetilde{\Gamma}_0(1;4,3) \vev{\frac{\delta f[A]}{\delta A(2)} \, G_A(3,4)} 
+ \widetilde{\Gamma}_0(2;4,3) \vev{\frac{\delta f[A]}{\delta A(1)} \, G_A(3,4)} \\
&+ \widetilde{\Gamma}_0(1;4,3) \, \widetilde{\Gamma}_0(2;6,5)
\vev{ f[A] \bigl[ G_A(3,4) \, G_A(5,6) - G_A(3,6) \, G_A(5,4) \bigr] } .
\end{split}
\ee
We stress that this is an exact identity, which holds for $f[A]$ being an arbitrary
functional of the gauge field only, i.e., not containing the momentum operator. Furthermore,
it will be sometimes convenient to express the last expectation value in terms of ghost fields
\be\label{tec3}
\vev{ f[A] \bigl[ G_A(3,4) \, G_A(5,6) - G_A(3,6) \, G_A(5,4) \bigr] } =
\vev{ f[A] \, c(3) \, \bar{c}(4) \, c(5) \, \bar{c}(6) } \, .
\ee


\subsection{\label{subsec:ke}Kinetic energy}
\begin{figure}
\includegraphics{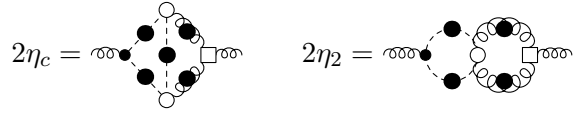}
\caption{\label{fig:etaterms}Diagrammatic representation of the contributions
\protect\eqref{etaterms} to the kinetic energy density.}
\end{figure}

After an integration by parts, the vacuum expectation value of the kinetic part of the
Yang--Mills Hamiltonian (first term in Eq.~\eqref{609-1}) can be expressed as
\be\label{ke1a}
E_k = \frac12 \int_\Omega \calD A \: \calJ[A] \:
\frac{\delta\psi[A]}{\delta A(1)} \: \frac{\delta\psi[A]}{\delta A(1)}
= \frac18 \vev{\frac{\delta S[A]}{\delta A(1)} \: \frac{\delta S[A]}{\delta A(1)}} .
\ee
The last expectation value has precisely the form of Eq.~\eqref{ke3} with $f[A]=1$
and the two external indices contracted. The terms in Eq.~(\ref{ke3}) involving
functional derivatives of $f[A]$ then vanish, and with the explicit form of the
action \eqref{ansatz} we find for the kinetic energy
\be\label{ke2}
E_k = \frac18 \left[ 2 \, \omega(1,1) + 2 \, \phi_t(1,1) +
\widetilde{\Gamma}_0(1;4,3) \, \widetilde{\Gamma}_0(1;6,5)
\vev{ c(3) \, \bar{c}(4) \, c(5) \, \bar{c}(6) } \right] .
\ee
Here, $\phi_t$ is the gluon tadpole term occurring in the gluon DSE (\ref{dse10}) and
being defined by Eq.~\eqref{dse10ae}. The ghost four-point function $\vev{c \bar{c} c \bar{c}}$
occurring in the last term can be expressed by means of Eq.~(\ref{cff4}) in terms of
propagators and proper functions. Contracting Eq.~\eqref{cff4} with the two bare
ghost-gluon vertices as in Eq.~\eqref{ke2}, one obtains
\be\label{ke2a}
\begin{split}
\widetilde{\Gamma}_0(1;4,3) \, & \widetilde{\Gamma}_0(1;6,5) \vev{ c(3) \, \bar{c}(4) \, c(5) \, \bar{c}(6) }
 = 4 \chi(1,3) D(3,4) \chi(4,1)
- \Bigl[
\widetilde{\Gamma}_0(1;5',6') \\
& +
\widetilde{\Gamma}_0(1;4,3)
\widetilde{\Gamma}(7;3',6')
\widetilde{\Gamma}(7';5',4')
G(3,3') G(4',4) D(7,7') \\
& +
\widetilde{\Gamma}_0(1;4,3)
\widetilde{\Gamma}(3',5',4',6') G(3,3') G(4',4)
\Bigr] G(6',6) \, G(5,5') \, \widetilde{\Gamma}_0(1;6,5) \, .
\end{split}
\ee
where $\chi(1,2)$ is the ghost loop defined by Eq.~\eqref{dse10ab}. The terms in the
square brackets represent precisely the first, third, and fifth term of the right-hand
side of the truncated DSE \eqref{ghdse5} for the ghost-gluon vertex. Therefore we can
use Eq.~\eqref{ghdse5} to rewrite the terms in the
bracket in Eq.~\eqref{ke2a} in more compact form
\be\label{ke3a}
E_k = \frac14 [ \omega(1,1) + \phi_t(1,1) - \chi(1,1)
+ 2 \chi(1,2) D(2,3) \chi(3,1) + \eta_c(1,1) - \eta_2(1,1) ] \, ,
\ee
where we have introduced the abbreviations
\be\label{etaterms}
\begin{split}
2 \eta_c(1,2) &=
\widetilde{\Gamma}_0(1;3,4) G(3',3)G(4,4')G(5,5')\widetilde{\Gamma}(6';5',3')
\widetilde{\Gamma}(7';4',5)D(6,6')D(7,7')\gamma_3(6,7,2) \, ,\\
2 \eta_2(1,2) &=
\widetilde{\Gamma}_0(1;3,4) G(3',3) G(4,4') \widetilde{\Gamma}(5',6';4',3') D(5,5') D(6,6') \gamma_3(5,6,2) \, ,
\end{split}
\ee
see Fig.~\ref{fig:etaterms}.
Eq.~\eqref{ke3a} can be slightly rewritten by using the gluon DSE \eqref{dse10} to
eliminate the variational kernel $\omega$ (or, more precisely, the sum $\omega+\phi_t$)
in favour of the inverse gluon propagator Eq.~\eqref{4-20}. Assuming furthermore
that the various loop terms Eqs.~\eqref{dse10a} are colour diagonal, e.g.\
\be
\label{ke4}
\xi^{ab}_{ij}(\vk) = \delta^{ab} \, t_{ij}(\vk) \, \xi(\vk) \, , \: \text{etc.}
\ee
which is guaranteed by global colour invariance, we can express the kinetic energy in
momentum space as
\be\label{ke5}
E_k = \frac{(N_c^2-1)(d-1)}{4} \: V \int \dbar{p}
\left\{ \frac{[\Omega(\vp)-\chi(\vp)]^2}{\Omega(\vp)} +
\xi(\vp) - \phi_1(\vp) + \phi_2(\vp) - 2 \eta_c(\vp) \right\} ,
\ee
where the terms $\phi_{1,2}$ are defined in Eqs.~\eqref{dse10ac} and \eqref{dse10ad},
and $\eta_c$ is given in Eq.~\eqref{etaterms}. Furthermore, in Eq.~\eqref{ke5} we have
introduced the abbreviation
\[
\dbar{p} \equiv \frac{\d[d]p}{(2\pi)^d} \, ,
\]
and $V$ is the spatial volume which arises as in Ref.~\cite{FeuRei04}.

The ghost and gluon loop contributions defined in Eqs.~\eqref{dse10a} read in momentum space
\begin{align}
\chi(\vp) &= \frac{t_{ij}(\vp)}{2 (N_c^2-1)(d-1)} \int \dbar{q}  \:
\widetilde{\Gamma}_{0,i}^{abc}(\vp;\vq-\vp,-\vq) \: \widetilde{\Gamma}_j^{acb}(-\vp;\vq,\vp-\vq)
G(\vq) G(\vp-\vq) \label{ke6a} \, , \\
\xi(\vp) &= \frac{1}{16 (N_c^2-1)(d-1)} \int \dbar{q} \: \d{k} \:
\frac{ ( \gamma_3 \circ \Gamma_3 )(\vp,\vq,\vk) }{\Omega(\vq) \, \Omega(\vk)} \:
\delta(\vp+\vq+\vk) \label{ke6} \, .
\end{align}
In the above equations, $\widetilde{\Gamma}$ and $\Gamma_3$ are, respectively, the full
ghost-gluon and three-gluon vertices defined by Eqs.~\eqref{263-1}, \eqref{269-1}.
We have also introduced here the contraction of two colour and Lorentz tensor structures
through transverse projectors as
\be
\label{ke7}
A \circ B \mathrel{\mathop:}= A^{abc}_{ijk}(\vp,\vq,\vk) \,
t_{il}(\vp) \, t_{jm}(\vq) \, t_{kn}(\vk) \,
B^{abc}_{lmn}(-\vp,-\vq,-\vk) .
\ee
Furthermore, the term $\eta_2$, Eq.~\eqref{etaterms}, has been discarded, since it
gives rise exclusively to higher-order contributions with more than three loops in the
energy, which are beyond our truncation scheme.


\subsection{\label{subsec:me}Magnetic energy}
\begin{figure}
\includegraphics{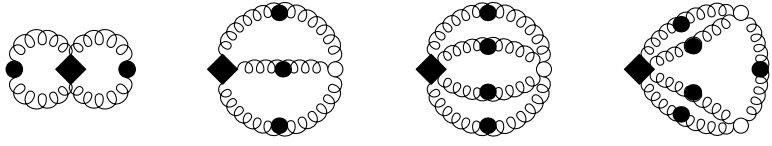}
\caption{\label{fig:magenergy}Contributions from the three- and four-gluon vertices
to the magnetic energy, from left to right: second term in Eq.~\protect\eqref{me1}, 
Eq.~\protect\eqref{me2}, Eq.~\protect\eqref{me3} and Eq.~\protect\eqref{me4}. The filled
diamonds stand for the bare vertices $T_3$, $T_4$ (Eqs.~\protect\eqref{maghamt})
occurring in the magnetic part of the Hamilton operator, Eq.~\protect\eqref{magham}.}
\end{figure}

In the notation of Eq.~(\ref{magham}) the magnetic energy is given by
\be\label{me1a}
E_B =  \frac{1}{4} \vev{ F^2_{ij} } =
-\frac12 \vev{A \partial^2 A} + \frac{g}{3!} \, T_3 \vev{A^3} + \frac{g^2}{4!} \, T_4 \vev{A^4} .
\ee
The first term on the right-hand side of Eq.~\eqref{me1a} can be expressed by means of
the gluon propagator \eqref{rrr}, while the second term can be expressed through the
proper three-point function $\Gamma_3$, Eq.~(\ref{tgv})
\be\label{me1b}
\frac{g}{3!} \, T_3 \vev{A^3} = - \frac{g}{3!} \, T_3 \circ \Gamma_3 \vev{AA}^3 .
\ee
In momentum space, these two energy contributions read
\be\label{me1}
\frac{(N_c^2-1)(d-1)}{4} \: V \int \dbar{p} \: \frac{\vp^2}{\Omega(\vp)}
- \frac{g \, V}{8 \cdot 3!} \int \dbar{p} \: \dbar{q} \: \d{k} \:
\frac{(T_3 \circ \Gamma_3 )(\vp,\vq,\vk)}{\Omega(\vp) \, \Omega(\vq) \, \Omega(\vk)} \: \delta(\vp+\vq+\vk) \, .
\ee
Let us turn now to the last term in Eq.~(\ref{me1a}), $\vev{A^4}$. The four-point
function is expressed by means of Eq.~\eqref{fgv} in terms of gluon propagators and
proper vertex functions. The disconnected terms in Eq.~\eqref{fgv}, i.e., the products
of two gluon propagators, when contracted with the bare four-gluon vertex $T_4$ \eqref{t4} results in
\be\label{me2}
g^2 \, \frac{N_c(N_c^2-1)}{16} \: V \int \dbar{p} \: \dbar{q} \:
\frac{d(d-3) + 3 -(\uvp\cdot\uvq)^2}{\Omega(\vp) \, \Omega(\vq)} \, .
\ee
which is the usual gluon tadpole term, which occurs already when a Gaussian wave
functional is used. Notice that, since the $q$ integral does not depend on an external
momentum, we can replace $\hat{q}_i \hat{q}_j \to \tfrac{1}{d} \delta_{ij}$ in the
integrand, and using
\[
d(d-3) +3 - \frac{1}{d} = \frac{(d-1)^3}{d}
\]
we can rewrite Eq.~(\ref{me2}) as
\be
\label{me2alt}
\tag{$\ref{me2}'$}
g^2 \, \frac{N_c(N_c^2-1)}{16} \: \frac{(d-1)^3}{d}\: V
\int \dbar{p} \: \dbar{q} \: \frac{1}{\Omega(\vp) \, \Omega(\vq)} \, .
\ee
Besides this, we get from the last term in Eq.~(\ref{me1a}) by using Eq.~\eqref{fgv}
also a contribution containing the proper four-point vertex function $\Gamma_4$
\be\label{me3}
- \frac{g^2 \, V}{16 \cdot 4!} \int \dbar{p} \: \dbar{q} \: \dbar{k} \: \d{\ell}
\frac{(T_4 \circ \Gamma_4)}{\Omega(\vp) \, \Omega(\vq) \, \Omega(\vk) \, \Omega(\vl)}
\: \delta(\vp+\vq+\vk+\vl) \, ,
\ee
and a contribution containing two three-gluon vertices
\be\label{me4}
\frac{g^2 \, V}{8} \int \dbar{[pqk\ell]}
\frac{T_{ijmn}^{abcd} \,  \Gamma^{abe}_{ijl}(-\vp,-\vq,\vp+\vq) \, \Gamma^{ecd}_{lmn}(\vk+\vl,-\vk,-\vl)}%
     {32 \Omega(\vp) \, \Omega(\vq) \, \Omega(\vk) \, \Omega(\vl) \, \Omega(\vp+\vq)} \,
(2\pi)^d \delta(\vp+\vq+\vk+\vl).
\ee
The Lorentz indices in Eq.~\eqref{me4} are supposed to be contracted by transverse
projectors, which we have not explicitly written down in order to prevent the equation
from getting cluttered. The various interaction contributions to the magnetic energy
given by Eqs.~\eqref{me1}--\eqref{me4} are shown in Fig.~\ref{fig:magenergy}.
Notice that the diagrams with a four-gluon vertex contain already three loops.


\subsection{\label{subsec:ce}Coulomb energy}
After an integration by parts, as in the case of the kinetic energy, the vacuum
expectation value of the Coulomb Hamiltonian, last term in Eq.~\eqref{609-1}, can be
expressed as
\be\label{coul1}
E_c = \frac{g^2}{2} \int \!\! \calD A \: \calJ[A] \int \!\!\d[d]x \, \d[d]y
\left[ \Hat{A}^{ac}_i(\vx) \, \frac{\delta \psi[A]}{\i \, \delta\!A_i^c(\vx)}\right]^*
F^{ab}_A(\vx,\vy) \left[ \Hat{A}^{bd}_j(\vy) \, \frac{\delta \psi[A]}{\i \, \delta\!A_j^d(\vy)}\right] .
\ee
In order to exploit our compact notation, we rewrite the colour charge density \eqref{615-1} as
\be\label{coul2}
\rho(1) = R(1;2,3) \, A(2) \: \frac{\delta}{\i \, \delta\!A(3)} \: ,
\ee
where
\be\label{coul2a}
R(1;2,3) = - R(1;3,2) = f^{a_1a_2a_3} \, \delta_{i_2i_3} \, \delta(\vx_1-\vx_2) \, \delta(\vx_1-\vx_3).
\ee
In this notation, the Coulomb energy \eqref{coul1} reads
\be\label{coul3}
E_c = \frac{g^2}{8} \: R(1;3,4) \, R(2;5,6)
\vev{F_A(1,2) \, A(3) \, A(5) \frac{\delta S[A]}{\delta A(4)} \: \frac{\delta S[A]}{\delta A(6)}}.
\ee
The remaining expectation value can, in principle, be expressed in terms of (so far
unknown) higher order vertex functions. The required manipulations are, however, quite
involved, and further simplifications are needed for practical reasons. Here we will again
restrict ourselves to terms containing up to three overlapping loops in the energy.
With this approximation we can factorize the Coulomb kernel $F_A$ (see also Ref.~\cite{FeuRei04}) as
\be\label{coul4}
E_c \simeq \frac{g^2}{8} \: R(1;3,4) \, R(2;5,6)
\vev{F_A(1,2)} \vev{ A(3) \, A(5) \, \frac{\delta S[A]}{\delta A(4)} \: \frac{\delta S[A]}{\delta A(6)}}.
\ee
The Coulomb propagator $\langle F_A \rangle$ is discussed later in Sec.~\ref{subsec:cff}.
The remaining expectation value has precisely the form (\ref{ke1}) with $f[A] = AA$,
and from Eq.~(\ref{ke3}) we obtain
\be\label{coul4a}
\begin{split}
&\biggl<\frac{\delta S[A]}{\delta A(4)} \: \frac{\delta S[A]}{\delta A(6)} \, A(3) \, A(5)  \biggr>=
\vev{\frac{\delta^2 S[A]}{\delta A(4) \, \delta A(6)} \, A(3) \, A(5)}
+\vev{\frac{\delta^2}{\delta A(4) \, \delta A(6)} [A(3) \, A(5)]} \\
& + \widetilde{\Gamma}_0(4;8,7) \vev{\frac{\delta}{\delta A(6)} [A(3) \, A(5)] \, G_A(7,8)} 
+ \widetilde{\Gamma}_0(6;8,7) \vev{\frac{\delta}{\delta A(4)} [A(3) \, A(5)]\, G_A(7,8)} \\
&+ \widetilde{\Gamma}_0(4;8,7) \, \widetilde{\Gamma}_0(6;8',7')
\vev{ A(3) \, A(5) \bigl[ G_A(7,8) \, G_A(7',8') - G_A(7,8') \, G_A(7',8) \bigr] } .
\end{split}
\ee
When inserted into Eq.~(\ref{coul4}), this expression can be simplified by exploiting
the colour antisymmetry of the vertices $R$, see Eq.~\eqref{coul2a}: For this reason,
the part of Eq.~\eqref{coul4a} symmetric with respect to the interchange of the indices
$(3 \leftrightarrow 4)$ or $(5 \leftrightarrow 6)$ vanishes. Furthermore, the two terms
on the r.h.s. of Eq.~\eqref{coul4a} with a single (bare) ghost-gluon vertex $\widetilde{\Gamma}_0$
yield identical contributions when inserted into Eq.~\eqref{coul4}.

The first four terms on the right-hand side of Eq.~\eqref{coul4a} can be straightforwardly
evaluated as in the preceding sections. With the explicit form of the ``action''
\eqref{ansatz}, one obtains for these terms
\be\label{coul4b}
\begin{split}
&{} t(3,6) \, t(4,5) + 2 \, \widetilde{\Gamma}_0(4;8,7) \, t(3,6) \vev{A(5) \, G_A(7,8)} 
+ 2 \, \omega(4,6) \vev{A(3) A(5)} +\\
&{} + \gamma(4,6,7) \vev{A(7) \,  A(3) \, A(5)} 
+ \tfrac12 \, \gamma(4,6,7,8) \vev{ A(7) \, A(8) \,  A(3) \, A(5)} =\\
&= t(3,6) \bigl[ t(4,5) - 4 \chi(4,7) D(7,5) \bigr] + 2 \omega(4,6) D(3,5) + \\
&{}- \gamma(4,6,7) \, D(5,5') \, D(3,3') \, D(5,5') \, \Gamma(3',5',7')
+ \tfrac12 \, \gamma(4,6,7,8) \vev{ A(7) A(8) A(3) A(5)} \, ,
\end{split}
\ee
where we have used Eqs.~\eqref{ggvdef}, \eqref{rrr}, and the definition of the ghost loop
Eq.~\eqref{dse10ab}. We are still left with the four-point function in Eq.~(\ref{coul4b}),
and with the six-point function in the last term in Eq.~(\ref{coul4a}). To work out
these terms, we notice that the Coulomb energy \eqref{coul4} can be diagrammatically
represented as
\[
\includegraphics{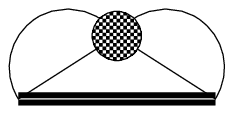}
\]
where the ``blob'' (including the four lines attached to it) represents the v.e.v.\
given  by the last bracket in Eq.~(\ref{coul4}) and the double line stands for the Coulomb propagator $\vev{F_A}$.
Since we need the energy up to three loops, we should keep only those contributions to
the ``blob'' which either factorize in two disconnected lines or which are either
irreducible or at most one-particle reducible (no box diagrams). At this order, for
the last term in Eq.~(\ref{coul4b}) it is sufficient to keep from the expression for
$\vev{A^4}$ given in Eq.~\eqref{fgv} only the disconnected terms, yielding
\be\label{coul4c}
\begin{split}
& \tfrac12 \, \gamma(4,6,7,8) \vev{ A(7) A(8) A(3) A(5)} = \\
&=\tfrac12 \, \gamma(4,6,7,8) \left[ D(3,5) \, D(7,8) +  D(3,7) \, D(5,8) + D(3,8) \, D(5,7) +\ldots \right]\\
&= D(3,5) \, 2 \, \phi_t(4,6) + \gamma(4,6,7,8) \, D(7,3) \, D(8,5) + \ldots
\end{split}
\ee
where we have used the definition of gluon tadpole $\phi_t$ (\ref{dse10ae}) and made
use of the symmetry properties of the four-gluon kernel $\gamma_4$. The dots in
Eq.~\eqref{coul4c} stand for terms which give rise to energy contributions with more
than three loops. Following the same line of reasoning, the last term in Eq.~\eqref{coul4a}
can be transformed to
\begin{align}
&\widetilde{\Gamma}_0(4;8,7) \, \widetilde{\Gamma}_0(6;8',7')
\vev{ A(3) \, A(5) \bigl[ G_A(7,8) \, G_A(7',8') - G_A(7,8') \, G_A(7',8) \bigr] } \nonumber \\
& \begin{aligned}
= \widetilde{\Gamma}_0(4;8,7) \, \widetilde{\Gamma}_0(6;8',7')
\bigl\{
D(3,5) & \vev{\bigl[ G_A(7,8) \, G_A(7',8') - G_A(7,8') \, G_A(7',8) \bigr]} \\
& + \vev{A(3) \, G_A(7',8')} \vev{A(5) \, G_A(7,8)} + \ldots\bigr\}
\end{aligned} \nonumber \\
& \begin{aligned}
= D(3,5) \bigl[ -2 \, \chi(4,6) &+ 4 \, \chi(4,7) D(7,8) \chi(8,6) + \ldots \bigr] \\
& + 4 D(3,7) \chi(7,6) D(5,8) \chi(8,4) + \ldots
\end{aligned} \label{coul4d}
\end{align}
where we used Eqs.~\eqref{ggvdef}, \eqref{cff4} and discarded terms involving more than
three loops in the energy. Collecting all terms given by Eqs.~(\ref{coul4a})--(\ref{coul4d})
and inserting the result into (\ref{coul4}), we finally obtain the Coulomb energy to
the desired (three-loop) order
\be\label{coul5a}
\begin{split}
E_c = \frac{g^2}{8} \: & R(1;3,4) \, R(2;5,6) \, F(1,2) \times \\
\times
\Bigl\{ {}&{}
2 \, D(3,5) \bigl[ \omega(4,6) + \phi_t(4,6) - \chi(4,6) + 2 \, \chi(4,7) \, D(7,8) \, \chi(8,6) \\
& + t(3,6) \bigl[ t(4,5) - 4 \chi(4,7) D(7,5) \bigr] + 4 D(3,7) \chi(7,6) D(5,8) \chi(8,4) \\
&- \gamma(4,6,7) \, D(5,5') \, D(3,3') \, D(5,5') \, \Gamma(3',5',7') \\
& + \gamma(4,6,7,8) \, D(7,3) \, D(8,5)
\Bigr\} .
\end{split}
\ee
where $F(1,2)=\vev{F_A(1,2)}$. For later use we rewrite this expression in momentum
space. Exploiting the symmetries of the entries and expressing the variational kernel
$\omega$ through the inverse gluon propagator $\Omega$, as we did for the kinetic energy,
Eq.~\eqref{coul5a} can be cast into the form
\be\label{coul5}
\begin{split}
E_c ={}& g^2 \, \frac{N_c(N_c^2-1)}{16} \: V \int \!\! \dbar{p} \, \dbar{q} \: F(\vp+\vq) \:
\frac{[d-2+(\uvp\cdot\uvq)^2]}{\Omega(\vp) \, \Omega(\vq)} \times \\
&{}\qquad \times \left\{ [\Omega(\vp)-\chi(\vp)-\Omega(\vq)+\chi(\vq)]^2 +
\xi(\vp) \, \Omega(\vp) + \xi(\vq) \, \Omega(\vq) \right\} \\
& - V \, \frac{g^2 }{8} \int \!\! \dbar{p} \, \dbar{q} \, \dbar{\ell} \: F(\vl)
\frac{t_{im}(\vp) \, t_{jn}(\vq) \, t_{kl}(\vp+\vq)}{\Omega(\vp) \, \Omega(\vq) \, \Omega(\vp+\vq)} \,
f^{gad} f^{gbe} \times\\
& \mspace{140mu}\times  \gamma^{cde}_{lmn}(\vp+\vq,\vl-\vp,-\vq-\vl) \, \Gamma^{cab}_{kij}(-\vp-\vq,\vp,\vq) \\
& + V \, \frac{g^2 }{8} \int \!\! \dbar{p} \, \dbar{q} \, \dbar{\ell} \: F(\vl)
\frac{t_{ij}(\vp) \, t_{lm}(\vq)}{\Omega(\vp) \, \Omega(\vq)} \, f^{abc} f^{ade}
\gamma^{bdce}_{iljm}(\vl-\vp,\vp,-\vl-\vq,\vq)
\end{split}
\ee
where $F(\vk)$ is the Fourier representation of the Coulomb propagator $\vev{F_A}$, see
Sec.~\ref{subsec:cff} below.
If we discard the three- and four-gluon kernels $\gamma_3$ and $\gamma_4$, which also
removes the gluon loop $\xi(\vp)$ Eq.~\eqref{dse10aa}, this expression reduces to the
Coulomb energy obtained in Refs.~\cite{FeuRei04,ReiFeu05} with a Gaussian wave functional,
\be\label{coul5b}
\begin{split}
E_c ={}& g^2 \, \frac{N_c(N_c^2-1)}{16} \: V \int \!\! \dbar{p} \, \dbar{q} \: F(\vp+\vq) \:
\frac{[d-2+(\uvp\cdot\uvq)^2]}{\Omega(\vp) \, \Omega(\vq)} \times \\
&{}\qquad \times \left\{ [\Omega(\vp)-\chi(\vp)-\Omega(\vq)+\chi(\vq)]^2 \right\} ,
\end{split}
\ee
The new features introduced by the inclusion of the three- and four-gluon kernels will
be studied in the subsequent sections.


\section{\label{sec:var}Determination of the variational kernels}

In the previous section we have expressed the vacuum energy $\bra{\psi} H \ket{\psi}$
in terms of the variational kernels $\omega$, $\gamma_3$, and $\gamma_4$ (occurring in our
ansatz \eqref{vac1}, \eqref{ansatz} for the vacuum wave functional $\psi[A]$) and of
the proper vertices $\Gamma_n$, $\widetilde{\Gamma}_n$. We now use the DSEs \eqref{dse10},
\eqref{dse11}, \eqref{dse12}, and \eqref{ghdse5} to express the proper vertex functions $\Gamma_n$,
$\widetilde{\Gamma}_n$ occurring in the energy in terms of the variational kernels
$\gamma_n$. We are then in a position to determine these kernels by minimizing
$\bra{\psi} H \ket{\psi}$. To make the calculations feasible, we will resort to a skeleton expansion
of $\bra{\psi} H \ket{\psi}$, keeping at most three-loop terms. As we will see, this is the
minimum number of loops required to obtain a non-trivial four-gluon kernel $\gamma_4$.
In the variation of the energy with respect to $\frac12\gamma_2=\omega$ and $\gamma_3$ we will
restrict ourselves up to two loops terms in $\bra{\psi} H \ket{\psi}$, which will be
sufficient to get a non-trivial $\gamma_3$ and a one loop gap equation for $\omega$.

\subsection{\label{subsec:xiphi}Three- and four-gluon kernel}
Below we determine the three-gluon kernel in leading order in the number of loops. For
this purpose, it is sufficient to keep up to two-loop terms in the energy. The relevant
contributions come then from the gluon
loop $\xi$ \eqref{ke6} occurring in the kinetic energy (\ref{ke5}), and the magnetic
energy contribution (\ref{me1}). These terms contain the three-gluon kernel $\gamma_3$
either explicitly or implicitly via the three-point proper vertex $\Gamma_3$, which
by its DSE (\ref{dse11}) is given in lowest order by the three-gluon kernel $\gamma_3$.
All remaining terms of $\Gamma_3$ contain additional loops and will henceforth be
discarded, resulting in the ``tree-level'' expression $\Gamma_3=\gamma_3$. Inserting
this expression in Eqs.~\eqref{ke6} and \eqref{me1} and taking into account the symmetry
of these kernels, the variation of these energy terms with respect to the three-gluon
kernel $\gamma_3$ leads to
\be\label{var1a}
\frac{\delta}{\delta\gamma_3}\int \dbar{p} \, \dbar{q} \, \dbar{\ell}
\frac{\delta(\vp+\vq+\vl)}{\Omega(\vp) \, \Omega(\vq) \, \Omega(\vl)}
\left[
(\gamma_3\circ\gamma_3) \frac{\Omega(\vp)+\Omega(\vq)+\Omega(\vl)}{4} - (\gamma_3\circ g T_3)
\right] \stackrel{!}{=} 0 \, ,
\ee
which fixes the three-gluon kernel to
\be\label{var1}
\gamma^{abc}_{ijk}(\vp,\vq,\vk) =
\frac{2 \, g \, T^{abc}_{ijk}(\vp,\vq,\vk)}{\Omega(\vp) + \Omega(\vq) + \Omega(\vk)} \, ,
\ee
where the tensor $T^{abc}_{ijk}(\vp,\vq,\vk)$ is defined in Eq. (\ref{t3}). The obtained
three-gluon kernel $\gamma_3$ is reminiscent of the perturbative one following from a
solution of the Yang--Mills Schr\"odinger equation in leading order in the coupling
constant $g$ \cite{Cam+09}
\be\label{var2}
\gamma^{(0)abc}_{ijk}(\vp,\vq,\vk) =
\frac{2 \, g \, T^{abc}_{ijk}(\vp,\vq,\vk)}{\abs{\vp} + \abs{\vq} + \abs{\vk}} \: , 
\ee
except that the perturbative gluon energy $\abs{\vk}$ is replaced in Eq.~\eqref{var1} by the non-perturbative
one $\Omega (\vk)$. Note that, in principle, the Lorentz indices in Eqs.~\eqref{var1}
and \eqref{var2} are contracted with transverse projectors, which we did not explicitly
write down, since they arise naturally as these kernels are always contracted with either
transverse gauge fields or the corresponding transverse Green functions.

The variational determination of the four-gluon kernel $\gamma_4$ is technically somewhat
more involved. The terms of $\vev{H}$ contributing to the variation with respect to
$\gamma_4$ are those containing $\gamma_4$ either explicitly or implicitly via the DSEs
\eqref{dse11}, \eqref{dse12} for the proper three- and four-gluon vertices $\Gamma_3$, $\Gamma_4$.
Terms explicitly containing $\gamma_4$ are the $\phi_{1,2}$ terms Eqs.~\eqref{dse10ac},
\eqref{dse10ad} of the kinetic energy Eq.~(\ref{ke5}), as well as the last term of the Coulomb energy \eqref{coul5}. Terms
containing $\Gamma_3$ or $\Gamma_4$ are given by the gluon loop $\xi$ \eqref{ke6} in
the kinetic energy \eqref{ke5} as well as by the magnetic energy contributions \eqref{me1}
and \eqref{me3}. All terms contributing to the variation of the energy with respect
to $\gamma_4$ are collected below:
\be\label{var3a}
\begin{split}
&\frac{1}{4! 2} \, \gamma(1,2,3,4) \, D(1,1') \,  D(2,2') \, D(3,3') \, \gamma(1',2',3',4) \\
&-\frac{1}{16} \, \gamma(1,2,3,4) \, \gamma(3',4',5) \, D(5,5') \, \gamma(5',1,2') \, D(2,2') \, D(3,3') \, D(4,4') +\\
&+\Gamma(1,2,3) \, D(1,1') \,  D(2,2') \left[ \frac{1}{16} \, \gamma(1',2',3) - \frac{1}{3!} \, D(3,3') \, T(1',2',3') \right] \\
&-\frac{1}{4!} \, \gamma(1,2,3,4) \, D(1,1') \,  D(2,2') \, D(3,3') \, D(4,4') \, \gamma(1',2',3',4')\\
&+ \frac{g^2}{8} \, F(1,2) \, R(1;3,4) \, R(2;5,6) \, \gamma(4,6,7,8) \, D(7,3) \, D(8,5) \, .
\end{split}
\ee
For simplicity, we have not explicitly symmetrized this expression with respect to a
permutation of the indices of $\gamma_4$.\footnote{%
Recall that, by definition, $\gamma_4$ is totally symmetric with respect to a permutation
of indices. Strictly speaking, one should first symmetrize Eq.~(\ref{var3a}) and then
take the variation. However, the same result is more conveniently obtained by taking
the variation of the unsymmetrized expression and symmetrizing afterwards.}
In the above expression we have used the DSE \eqref{dse12} to replace the four-gluon
vertex $\Gamma_4$ by the four-gluon kernel $\gamma_4$. To be consistent, we have to
use the DSE for the three-gluon vertex, Eq.~(\ref{dse11}), to express the vertex function
$\Gamma_3$ in \eqref{var3a} by the variational kernels, where it is sufficient to
retain only terms involving $\gamma_4$ or $\Gamma_4$, and the latter is to be replaced
by $\gamma_4$ due to the DSE \eqref{dse12}. Taking into account the symmetry properties
of the quantities involved, by Eq.~\eqref{dse11} we are led to make the following
replacement in Eq.~(\ref{var3a})
\be\label{var3b}
\begin{split}
\Gamma(1,2,3) \to {}& - \frac12 \, \gamma(3,4,5) \, D(4,4') \, D(5,5') \, \gamma(4',5',1,2) \\
&\qquad -\gamma(3,1,4,5) \, D(4,4') \, D(5,5') \, \gamma(4',5',2) \, .
\end{split}
\ee
With this replacement, the variation of Eq.~\eqref{var3a} with respect to $\gamma_4$
is now straightforward and yields after proper symmetrization with respect to external
indices of $\gamma_4$
\be\label{var3}
\begin{split}
\bigl[ \Omega(\vk_1) &+ \Omega(\vk_2) + \Omega(\vk_3) + \Omega(\vk_4) \bigr] \, \gamma^{abcd}_{ijkl}(\vk_1,\vk_2,\vk_3,\vk_4) =
2 \, g^2 \, T^{abcd}_{ijkl} \\
-\frac12 &\biggl\{
\gamma^{abe}_{ijm}(\vk_1,\vk_2,-\vk_1-\vk_2) \, t_{mn}(\vk_1+\vk_2) \, \gamma^{cde}_{kln}(\vk_3,\vk_4,\vk_1+\vk_2) \\
&{}\qquad + \gamma^{ace}_{ikm}(\vk_1,\vk_3,-\vk_1-\vk_3) \, t_{mn}(\vk_1+\vk_3) \, \gamma^{bde}_{jln}(\vk_2,\vk_4,\vk_1+\vk_3) \\
&{}\qquad\qquad + \gamma^{ade}_{ilm}(\vk_1,\vk_4,-\vk_1-\vk_4) \, t_{mn}(\vk_1+\vk_4) \gamma^{bce}_{jkn}(\vk_2,\vk_3,\vk_1+\vk_4)
\biggr\} \\
-2 g^2 & \biggl\{
f^{abe} f^{cde} \delta_{ij} \delta_{kl}
\bigl[\Omega(\vk_1) - \Omega(\vk_2)\bigr] F(\vk_1+\vk_2) \bigl[\Omega(\vk_3) - \Omega(\vk_4)\bigr] \\
&{}\qquad + f^{ace} f^{bde} \delta_{ik} \delta_{jl}
\bigl[\Omega(\vk_1) - \Omega(\vk_3)\bigr] F(\vk_1+\vk_3) \bigl[\Omega(\vk_2) - \Omega(\vk_4)\bigr] \\
&{}\qquad\qquad + f^{ade} f^{bce} \delta_{il} \delta_{jk}
\bigl[\Omega(\vk_1) - \Omega(\vk_4)\bigr] F(\vk_1+\vk_4) \bigl[\Omega(\vk_2) - \Omega(\vk_3)\bigr]
\biggr\}.
\end{split}
\ee
$F(\vp)$ is again the Fourier representation of the Coulomb propagator $\vev{F_A}$.
Eq.~(\ref{var3}) yields a four gluon kernel $\gamma_4$, which is reminiscent of the
perturbative one \cite{Cam+09} except that the perturbative propagators and vertices
are replaced by the full ones.


\subsection{\label{subsec:gapeq}Gap equation}
Given the explicit form of the energy functional \eqref{ke5}, \eqref{me1}--\eqref{me4},
\eqref{coul5}, it is more convenient to use the DSE \eqref{dse10}
for the gluon propagator to express the kernel $\omega(\vp)$ in terms of the gluon
energy $\Omega(\vp)$, and vary the energy density with respect to $\Omega^{-1}(\vp)$. The vacuum energy is given
by closed loop diagrams, and the variation with respect to the gluon propagator $\Omega^{-1}(\vp)$ reduces the number
of loops by one. If the (gap) equation for $\Omega(\vp)$ is to be calculated up to one loop, it is
sufficient to keep up to two loops in the energy. At this order, the energy density
$\eps$ defined by $\vev{H} =\mathrel{\mathord:} V (d - 1) (N^2_c - 1) \eps $ is given by
\be\label{ge1}
\begin{split}
\eps &= \frac14 \int \dbar{p} \: \frac{\vp^2+[\Omega(\vp) - \chi(\vp)]^2}{\Omega(\vp)} \\
&- \frac{g^2 \, N_c}{3! 8 (d-1)} \int \dbar{p} \dbar{q} 
\frac{\gamma_3}{\Omega(\vp)\Omega(\vq)\Omega(\vp+\vq)} \circ
\left[ \frac{\Omega(\vp)+\Omega(\vq)+\Omega(\vp+\vq)}{4} \, \gamma_3 - g T_3 \right] \\
&+ \frac{g^2 \, N_c}{16(d-1)} \int \dbar{p} \: \dbar{q} \: [d-2+(\uvp\cdot\uvq)^2] F(\vp+\vq)
\frac{[\Omega(\vp)-\chi(\vp)-\Omega(\vq)+\chi(\vq)]^2}{\Omega(\vp) \, \Omega(\vq)} \, ,
\end{split}
\ee
see Eqs.~\eqref{ke5}, \eqref{me1} and \eqref{coul5b}. In Eq.~(\ref{ge1}) we have
discarded the tadpole term, Eq.~\eqref{me2} or \eqref{me2alt}, since it represents an
irrelevant constant, which disappears after renormalization. Except for the gluon loop
(second term), the energy density Eq.~(\ref{ge1}) was already obtained in
Ref.~\cite{FeuRei04,ReiFeu05}, where a Gaussian wave functional multiplied by
$\calJ^{-1/2}[A]$ was used. The gluon loop is lost when a Gaussian wave functional is
used, for which Green's functions with an odd number of fields vanish.

In principle, in Eq.~\eqref{ge1} the three-gluon kernel $\gamma_3$ \eqref{var1} obtained
from the variational principle $\delta\epsilon/\delta\gamma_3 = 0$ depends on $\Omega(\vp)$.
However, since the energy density $\eps$ \eqref{ge1} with $\gamma_3$ given by Eq.~\eqref{var1}
is already stationary with respect to variations of $\gamma_3$, we can ignore the implicit
$\Omega(\vp)$ dependence of $\eps$ via $\gamma_3$. Variation of $\eps$ \eqref{ge1} with
respect to $\Omega^{-1}(\vk)$ then yields the gap equation 
\be\label{ge2}
\Omega(\vp)^2 = \vp^2 + \chi(\vp)^2 - I_\mathrm{G}(\vp) + I_\mathrm{C}(\vp),
\ee
where $\chi(\vp)$ is the ghost loop \eqref{ke6a},
\begin{subequations}\label{ge3}
\be\label{ge3a}
I_\mathrm{G}(\vp) = \frac{1}{4(d-1)(N_c^2-1)} \int \frac{\d[d]q}{(2\pi)^d} \:
\frac{1}{\Omega(\vq)\,\Omega(\vp+\vq)} \, \gamma_3 \circ
\left[ g T_3 - \gamma_3 \, \frac{\Omega(\vp)+\Omega(\vp+\vq)}{4} \right]
\ee
is the gluon loop contribution, and
\be\label{ge3b}
\begin{split}
I_\mathrm{C}(\vp) = \frac{g^2 \, N_c}{2(d-1)} \int \frac{\d[d]q}{(2\pi)^d} \:
& \bigl[ d-2+(\uvk\cdot\uvq)^2 \bigr]  \, F(\vp+\vq) \\
& \times \frac{\bigl[\Omega(\vq)-\chi(\vq)+\chi(\vp)\bigr]^2-\Omega(\vp)^2}{\Omega(\vq)}
\end{split}
\ee
\end{subequations}
arises from the Coulomb term. Except for the gluon loop $I_\mathrm{G}(\vk)$, Eq.~(\ref{ge2})
is the gap equation found already in \cite{FeuRei04}.

If we insert the expression derived above in leading order for the three-gluon kernel
$\gamma_3$, Eq.~\eqref{var1}, into the gluon-loop contribution Eq.~\eqref{ge3a}, this
takes the form
\be\label{gl-loop-expl}
I_\mathrm{G}(\vp) = \frac{g^2 \, N_c}{d-1} \int \frac{\d[d]q}{(2\pi)^d} \:
\frac{2\Omega(\vp)+\Omega(\vq)+\Omega(\vp+\vq)}{\bigl[\Omega(\vp)+\Omega(\vq)+\Omega(\vp+\vq)\bigr]^2} \:
\frac{\Sigma(\vp,\vq)}{\Omega(\vq)\,\Omega(\vp+\vq)} \, ,
\ee
where the function $\Sigma$ arises from the contraction of the Lorentz structure of
two three-gluon vertices upon imposing momentum conservation
\be
\label{1183-1}
T_3 \circ T_3 \bigr|_{\vk=-\vp-\vq} =\mathrel{\mathop:} 4 \, N_c \, (N_c^2-1) \, \Sigma(\vp,\vq) \, ,
\ee
and is given explicitly by
\be\label{ge1a}
\Sigma(\vp,\vq) = \bigl[ 1-(\uvp\cdot\uvq)^2 \bigr]
\left[ (d-1)(\vp^2+\vq^2) + \frac{(d-2)\vp^2\vq^2+(\vp\cdot\vq)^2}{(\vp+\vq)^2} \right].
\ee

To exhibit the UV-behaviour of the various loop terms, let us consider them in leading order in
perturbation theory \cite{CamReiWeb09,Cam+09}, where $F(\vp)=1/\vp^2$ and $\chi(\vp)=0$.
In this order we find with 3-momentum cut-off $\Lambda$
\begin{subequations}\label{ge5}
\begin{align}
I_\mathrm{G}(\vp) &= \frac{g^2 N_c}{(4\pi)^2}
\left[ \frac{4}{3} \: \Lambda^2 + \frac{22}{15} \: \vp^2 \ln\frac{\Lambda^2}{\vp^2} \right]\label{ge5a} , \\
I_\mathrm{C}(\vp) &= \frac{g^2 N_c}{(4\pi)^2}
\left[ \frac{4}{3} \: \Lambda^2 - \frac{8}{15} \: \vp^2 \ln\frac{\Lambda^2}{\vp^2} \right]\label{ge5b}.
\end{align}
\end{subequations}
One observes that the quadratic divergence in Eq.~\eqref{ge2} cancels, and the sum of
the logarithmic ones is consistent with the result of Lagrangian-based perturbation
theory in Coulomb gauge \cite{WatRei07a,WatRei07b}
\be
\frac{\abs{\vp}}{\Omega(\vp)} = 1 + \frac{1}{2\vp^2} [ I_\mathrm{G}(\vp) - I_\mathrm{C}(\vp) ] ) + \ldots =
1 + \frac{g^2 N_c}{(4\pi)^{2-\eps}} \left( \frac1\eps - \ln\frac{\vp^2}{\mu^2} \right) + \ldots
\ee
where we have used dimensional regularization with $d=3-2\eps$.

\begin{figure}
\includegraphics[width=.45\linewidth]{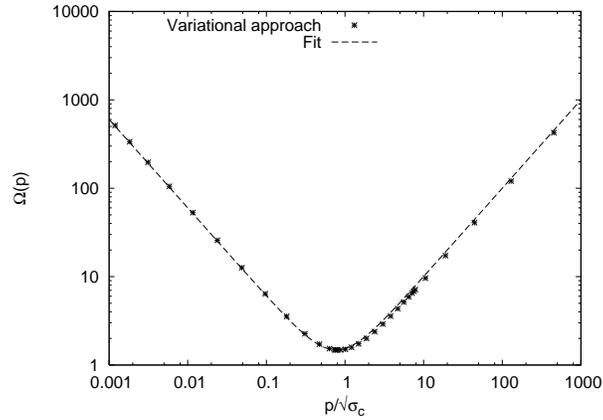}
\caption{\label{fig:omega}Numerical results from Ref.~\protect\cite{EppReiSch07} for the
gluon energy $\Omega(\vp)$ and fit to Eq.~\protect\eqref{ge3a}.}
\end{figure}

\begin{figure}
\centering
\includegraphics[width=.45\linewidth]{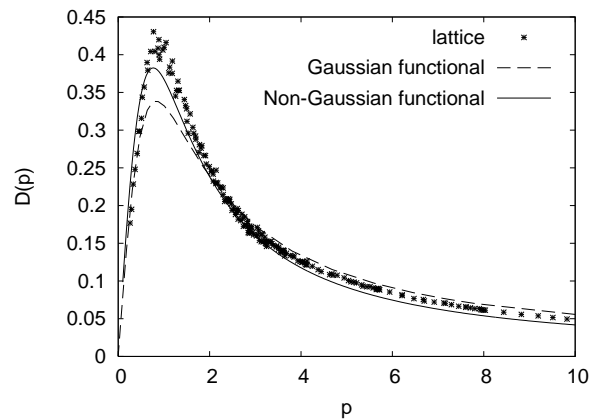}
\caption{\label{fig:gluonloop}Comparison of the gluon propagator obtained in
Ref.~\protect\cite{EppReiSch07} with a Gaussian wave functional (dashed line) with the
corrected one from Eq.~\protect\eqref{ge6} (full line), and the lattice data from
Ref.~\protect\cite{BurQuaRei09}.}
\end{figure}

To estimate the size of the contribution of the gluon loop \eqref{ge3a} to the gap
equation \eqref{ge2}, we use the gluon energy $\Omega$ obtained with a Gaussian wave 
functional in Ref.~\cite{EppReiSch07} as input. For practical purposes, we fit the
$\Omega(\vp)$ obtained in Ref.~\cite{EppReiSch07} to the formula
\be\label{gribov}
\Omega(\vp) = \sqrt{ \vp^2 + \frac{m_\textsc{a}^4}{\vp^2} + c^2} \, ,
\ee
which yields $m_\textsc{a}^4 \simeq 0.36 \sigma_\textsc{c}^2$, $c^2 \simeq 1.0 \sigma_\textsc{c}$.
The numerical results of Ref.~\cite{EppReiSch07} and the fit to Eq.~\eqref{gribov} are shown in Fig.~\ref{fig:omega}.
The gluon loop \eqref{ge3a} is UV divergent. In principle, the UV-divergent part is
removed by the renormalization of the gap equation \eqref{ge2}, which can be done
analogously to Refs.~\cite{ReiEpp07,Epp+07}.
We therefore calculate
\be\label{ge6}
\Omega_{\mathit{new}}(\vp)^2 = \Omega_{\mathit{old}}(\vp)^2 -
( I_\mathrm{G}(\vp)[\Omega_{\mathit{old}}] - I_\mathrm{G}(\vp)_{\mathit{div}} ) \, ,
\ee
where $I_\mathrm{G}(\vp)_{\mathit{div}}$ is the (known) perturbative divergent part of
the gluon loop, which has been subtracted, and Eq.~\eqref{gribov} has been used for
$\Omega_{\mathit{old}}(\vp)$. The (inverse) gluon propagator $\Omega_{\mathit{new}}(\vp)$
Eq.~\eqref{ge6} is shown in Fig.~\ref{fig:gluonloop} together with the one obtained
previously \cite{EppReiSch07} with a gaussian wave functional $\Omega_{\mathit{old}}(\vp)$.
The mismatch in the UV is due to the anomalous dimension developed by $\Omega_{\mathit{new}}(\vp)$
and absent in $\Omega_{\mathit{old}}(\vp)$. As seen in Fig.~\ref{fig:gluonloop}, significant
correction from the gluon loop arises in the mid-momentum regime, a behaviour which was
observed also in Landau gauge \cite{LanReiGat01,FisMaaPaw09}. This a posteriori supports
the use of the Gaussian wave functional for the description of the infrared regime.


\subsection{\label{subsec:cff}The Coulomb form factor}

With the explicit expression of the three- and four-gluon kernels, $\gamma_3$ Eq.~\eqref{var1}
and $\gamma_4$ Eq.~\eqref{var3}, on hand, we are left with three coupled equations: Eq.~\eqref{ghdsemom}
for the ghost propagator, Eq.~\eqref{ghdse5} for the ghost-gluon vertex, and the gap equation \eqref{ge2}
for the gluon propagator. The final piece which is missing for closing this set of equations is the non-Abelian
colour Coulomb potential $F(1,2)$ defined by the
vacuum expectation value of the Coulomb kernel Eq.~\eqref{620-1}
\be\label{cff2}
F(1,4) = \vev{F_A(1,4)} = \vev{ G_A(1,2) \, G_0^{-1}(2,3) \, G_A(3,4) } .
\ee
The quantity $g^2 F(1,2)$ directly relates to the heavy quark potential \cite{PopWatRei10} and is hence
a renormalization group invariant quantity. Before we come to the evaluation of $\vev{F_A}$, let us
remark that in practical application it is not necessary to explicitly solve the DSE \eqref{ghdse5}
for the ghost-gluon vertex. Rather it is sufficient to replace the full ghost-gluon vertex $\widetilde{\Gamma}$
Eq.~\eqref{263-1} by the bare one $\widetilde{\Gamma}_0$ Eq.~\eqref{5-16}. This approximation
is motivated by the ``non-renormalization'' theorem for this vertex \cite{Tay71}.
Although this theorem was originally proven \cite{Tay71} and confirmed on the lattice
\cite{CucMenMih04,Ste+06} for QCD in Landau gauge, the arguments carry over to the
present case of Coulomb gauge. A perturbative evaluation of the ghost-gluon vertex in
Coulomb gauge shows indeed that its quantum corrections are finite and independent of
the scale \cite{CamReiWeb09,Cam+09}.

The vacuum expectation value $\vev{F_A}$ is commonly expressed in terms of the Coulomb
form factor $f$. This quantity measures the deviation of $\vev{F_A}$ from the factorized
form $\vev{G_A} G_0^{-1} \vev{G_A}$ and is defined in momentum space by
\be\label{cff7a}
F(\vp) =\mathrel{\mathop:} G(\vp) \, f(\vp) \, \vp^2 \, G(\vp) \, ,
\ee
where $G=\vev{G_A}$ is the ghost propagator \eqref{246-1}. By taking the vacuum expectation
value of the operator identity
\be\label{cff10}
F_A = \frac{\partial}{\partial g} \bigl( g \, G_A \bigr) \, ,
\ee
the Coulomb form factor $f(\vp)$ can be related to the ghost form factor $d(\vp)$ and
from the ghost DSE \eqref{ghdsemom} the following (approximate) integral equation for
the Coulomb form factor is obtained \cite{FeuRei04,Swi88}
\be\label{cff8b}
f(\vp) = 1 + g^2 \frac{N_c}{2} \int \dbar{q} \: \frac{1-(\uvp\cdot\uvq)^2}{\Omega(\vq)}  \:
(\vp-\vq)^2 \, G^2(\vp-\vq) \, f(\vp-\vq) \, .
\ee
In the derivation of this equation no assumption on the form of the vacuum wave functional
enters, so this equation remains also valid for non-Gaussian wave functionals.

With the approximation $\widetilde{\Gamma}$, Eq.~\eqref{263-1}, $\to\widetilde{\Gamma}_0$, 
Eq.~\eqref{5-16}, Eqs.~\eqref{ghdsemom}, \eqref{ge2}, and \eqref{cff8b} form a closed set of coupled
equations for the ghost propagator $G(\vp)$, the gluon energy $\Omega(\vp)$, and the
Coulomb form factor $f(\vp)$, whose solutions provide the variational solution of the
Yang--Mills Schr\"odinger equation. These equations have to be renormalized, which can be done in exactly
the same way as in Refs.~\cite{FeuRei04,ReiEpp07,Epp+07}. The numerical solution of
this set of equations can be carried out as in the case of the Gaussian wave functional
\cite{FeuRei04,EppReiSch07}and is subject to future work. 

As already mentioned before, the use of a Gaussian wave functional misses the contribution
of the bare three-gluon vertex and thus the gluon loop in the gap equation. Fortunately,
as we have seen in the previous subsection,
the gluon loop is IR subleading compared to the (included) ghost loop, and thus irrelevant
for the IR behaviour of the Green functions. Therefore the results of Refs.~\cite{FeuRei04,EppReiSch07}
remain fully valid in the IR. In particular, the gluon propagator does not change in
the IR. The gluon loop does, however, matter in the UV and is responsible for the
anomalous dimension of the gluon propagator, which enters the running coupling constant.


\section{\label{sec:vertices}The three- and four-gluon vertex}

As we have seen, the gap equation differs from the one obtained in \cite{FeuRei04} only
by the gluon loop contribution, which gives sizeable corrections in the mid-momentum
regime. In the present section, we investigate the three- and four-gluon vertices $\Gamma_{3,4}$ by
using the ghost and gluon propagators determined with a Gaussian wave functional
\cite{EppReiSch07} as input. We will not resort to the tree-level result $\Gamma_n=\gamma_n$
but rather solve the corresponding DSEs to one-loop order.

\subsection{Solution of the truncated three-gluon vertex DSE}
The DSE for the three gluon-vertex at one-loop order is given by Eq.~\eqref{dse11}.
Assuming ghost dominance (see for example Ref.~\cite{SchLedRei06}), we keep only the
ghost-loop (the third term on the right-hand side of Eq.~\eqref{dse11} and Fig.~\ref{fig:3gvDSE}).
Furthermore, we replace the full ghost-gluon vertex by the bare one, see Sec.~\ref{subsec:cff}.
After extracting the colour structure, the truncated DSE for the three-gluon vertex becomes
\be\label{3gvff1a}
\Gamma_{ijk}(\vp,\vq,\vk) = \gamma_{ijk}(\vp,\vq,\vk) 
 + \i g^3 N_c \int \dbar\ell \: G(\vl) \, G(\vl-\vp) \, G(\vl+\vq) \ell_i \ell_j (\ell-p)_k \, ,
\ee
where momentum conservation $\vk+\vp+\vq=0$ is implied and we have omitted longitudinal
terms, which, by definition, cannot enter the proper $n$-point vertex functions \eqref{269-1}
of the transverse (Coulomb) gauge field. Possible tensor decompositions of the three-gluon
proper vertex are given in Refs.~\cite{BalChi80b,AlkHubSch09}. Here, for sake of illustration, we
confine ourselves to the form factor corresponding to the tensor structure of the bare
three-gluon vertex, defined by
\be\label{3gvff1}
f_{3A} \mathrel{\mathop:}= \frac{\Gamma^{(0)}_3 \circ \Gamma_3}{\Gamma^{(0)}_3 \circ \Gamma^{(0)}_3} \, ,
\ee
where $\Gamma^{(0)}_3$ is the perturbative vertex given in Eq.~\eqref{var2}. Restricting
the kinematic configuration to the case where two external momenta have the same magnitude, i.e.,
\be
 \vq^2 = \vp^2 = p^2 , \qquad \vq\cdot\vp = c p^2 \, ,
\ee
the form factor Eq.~\eqref{3gvff1} depends only on two variables, the magnitude $p$
of the two momenta and the cosine $c$ of the angle between them. Contracting
Eq.~\eqref{3gvff1a} with the bare three-gluon vertex Eq.~\eqref{var2}, thereby using the
variational kernel $\gamma_3$, Eq.~\eqref{var1}, and the tensor structure $T_3$ given in
Eq.~\eqref{t3}, yields the following equation for the form factor $f_{3A}$ Eq.~\eqref{3gvff1}
\be\label{3gvff2}
\begin{split}
f_{3A}(p^2,c) ={} &
\frac{2p + \sqrt{2p^2(1+c)}}{2\Omega(p^2)+\Omega(2p^2(1+c))} \\
&- g^2 \frac{N_c}{4} \: \frac{2 + \sqrt{2 (1+c)}}{p (1-c) (9+8c+c^2)}
\int \dbar\ell \: G(\vl) \: G(\vl-\vp) \: G(\vl+\vq) \: \mathcal{B}(\vl,\vp,\vq) \, ,
\end{split}
\ee
where we have introduced the abbreviation
\be\label{3gvff3}
\begin{split}
\mathcal{B}(\vl,\vp,\vq) ={}& \vl^2 \bigl[ p^2(1-c)-(1+2c)(\vl\cdot\vq-\vl\cdot\vp) \bigr]
 + \frac{(\vl\cdot\vq)^3}{p^2} - \frac{(\vl\cdot\vp)^3}{p^2} \\
& - (1+c) \bigl[ (\vl\cdot\vq)^2 + (\vl\cdot\vp)^2 \bigr] \\
& + (\vl\cdot\vq)(\vl\cdot\vp) \bigl[ c(1+c)+2+\frac{1-c}{p^2} (\vl\cdot\vq-\vl\cdot\vp) \bigr] .
\end{split}
\ee
The apparent singularity at $c=1$ in the coefficient in front of the integral in
Eq.~\eqref{3gvff2} is cancelled by the numerator, and the whole term is regular in the
collinear limit. Also the singularities at $\vl=0$, $\vl=\vp$, and $\vl=-\vq$ are integrable.

\begin{figure}
\includegraphics[width=.5\linewidth]{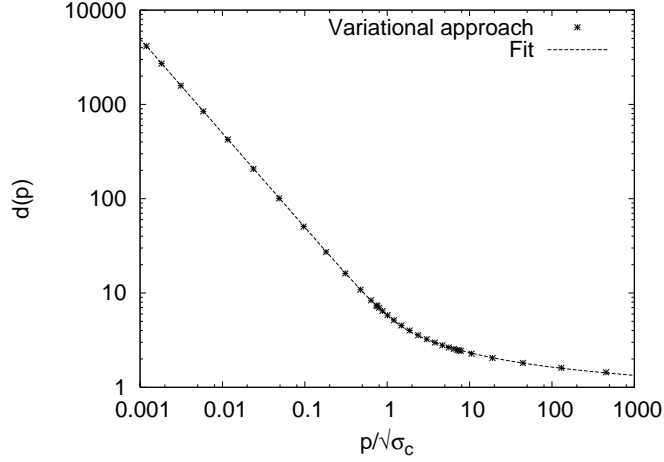}
\caption{\label{fig:ghost} Ghost form factor from \protect\cite{EppReiSch07} and fit
to Eq.~\protect\eqref{ghostfit}.}
\end{figure}

For the actual calculation of $f_{3A}$, Eq.~\eqref{3gvff1}, we use the numerical result
for $\Omega(\vp)$ and $G(\vp)$ obtained in Ref.~\cite{EppReiSch07} (with a Gaussian wave functional)
as input. For $\Omega(\vp)$ we use the parametrization Eq.~\eqref{gribov} while the
ghost form factor $d(\vp)=\vp^2 G(\vp)$ is parametrized as
\be\label{ghostfit}
d(x) = a \sqrt{ \frac{1}{x^2} + \frac{1}{\ln(x^2+c^2)} } \, , \quad
x^2 = \frac{\vp^2}{\sigma_\textsc{c}} \, ,
\ee
with $\sigma_\textsc{c}$ being the Coulomb string tension.The fit to the data shown in
Fig.~\ref{fig:ghost} yields $a\simeq5$ and $c\simeq4$.

The form \eqref{ghostfit} of the ghost form factor embodies also the
non-perturbative anomalous dimension, which in turn guarantees the convergence of the
integral in Eq.~\eqref{3gvff2}. 

\begin{figure}
\includegraphics[width=.45\linewidth]{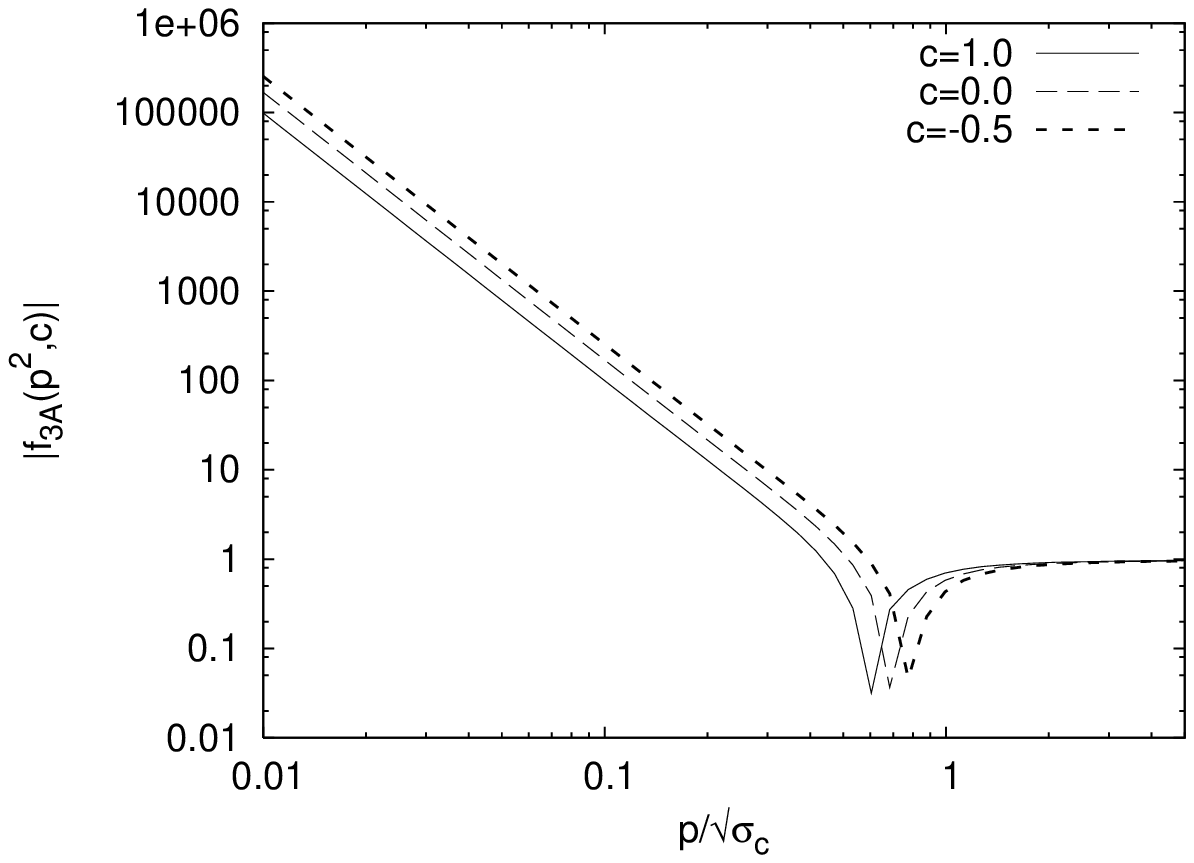} \:
\includegraphics[width=.45\linewidth]{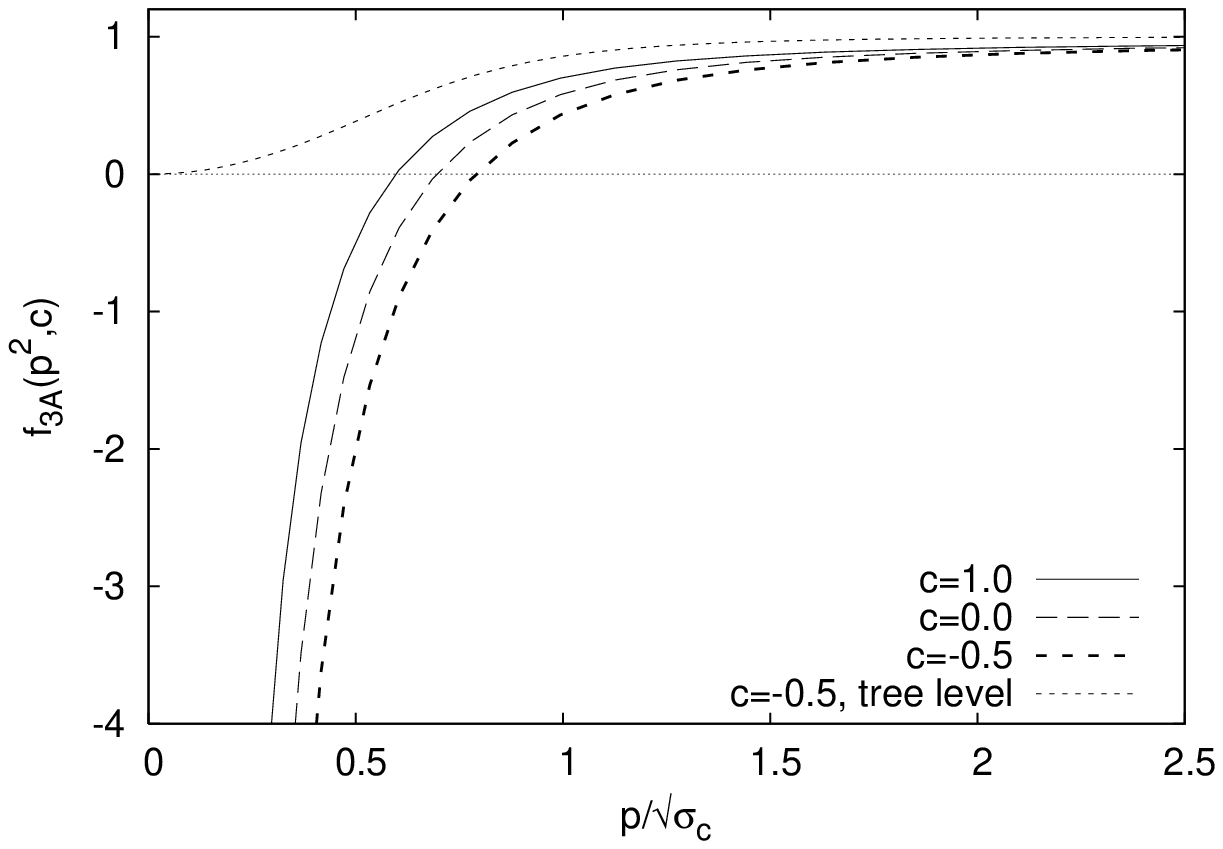}
\caption{\protect\label{fig:3gv} Form factor $f_{3A}(p^2,c)$ defined in Eq.~\eqref{3gvff1}
for two momenta of equal magnitude $p$ plotted for various values of the relative angle $c$.}
\end{figure}

The numerical results for $f_{3A}(p^2,c)$ are shown in Fig.~\ref{fig:3gv} for some
values of the cosine of the relative angle, $c$.
The logarithmic plot (left panel) shows that the curves for different $c$ have the same
power law behaviour in the infrared region, namely $p^{-3}$, in agreement with the IR
analysis of the ghost loop carried out in Ref.~\cite{AlkHubSch09}, according to which
the IR exponent of the form factor $f_{3A}$ should be three times the one of the ghost
dressing function Eq.~\eqref{ghostfit}.\footnote{%
In Ref.~\cite{SchLedRei06}, due to the use of a Gaussian wave functional, the bare term
escaped, and only the ghost loop was calculated for a different solution of the DSEs
\cite{FeuRei04}, which leads to less IR singular ghost and gluon propagators.}
The linear plot (right panel) shows that the form factor approaches
unity in the high-momentum regime and changes sign in the mid momentum regime. 

\begin{figure}
\includegraphics[width=.45\linewidth]{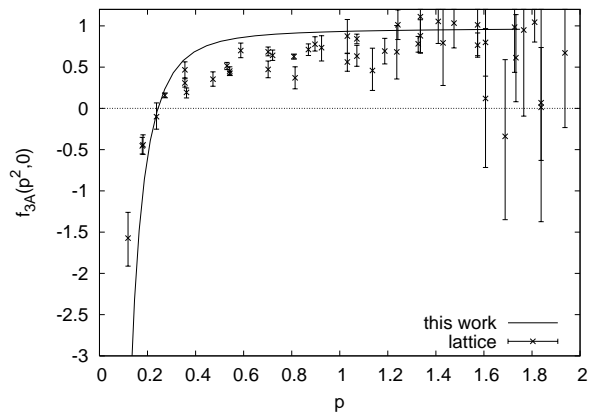}
\caption{\protect\label{fig:3gv-lat} Form factor $f_{3A}$ of the three-gluon
vertex for orthogonal momenta and comparison to lattice data for the $d=3$ Landau-gauge
vertex \cite{CucMaaMen08}. The momentum scale is arbitrary and has been adjusted to
make the sign change occur at the same point. The lattice data are shown by courtesy of A.~Maas.}
\end{figure}

In Ref.~\cite{CucMaaMen08} the form factor $f_{3A}$ Eq.~\eqref{3gvff1} of the three-gluon
vertex was evaluated on the lattice for $d=3$ Yang--Mills theory in Landau gauge. The
result is shown in Fig.~\ref{fig:3gv-lat} and compares well to our result in low-momentum
regime.
In particular, in both studies, the sign change of the form factor occurs roughly at
the same momentum where the gluon propagator has its maximum. The $d=3$ Yang--Mills
theory in Landau gauge can be interpreted as using the wave functional
\be
\psi[A] = \mathcal{N} \, \exp\left[ - \frac{1}{4\mu^2} \int (F_{ij}^a)^2 \right]
\ee
in the Hamiltonian approach to $d=3+1$ Yang--Mills theory in Coulomb gauge. This wave
functional can be considered to represent the strong coupling limit of the true vacuum
wave functional \cite{Gre79+87}. In Ref.~\cite{QuaReiBur10} it was shown by means
of lattice calculations that this wave functional yields static propagators which in
the IR compare well with those obtained in $d=3+1$ Yang--Mills theory in Coulomb gauge.
Thus we can conclude from Fig.~\ref{fig:3gv-lat} that our results compare favourably
with the lattice data.

\subsection{Estimate of the four-gluon vertex}
\begin{figure}
\includegraphics[width=.45\linewidth]{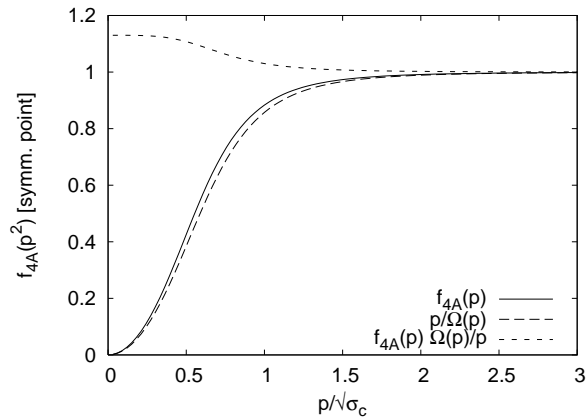}
\caption{\protect\label{fig:4gv} Form factor $f_{4A}(p^2)$ of the four-gluon
vertex at the symmetric point. The leading contribution $p/\Omega(p)$ is also shown.}
\end{figure}

Due to its DSE \eqref{dse12}, the four-gluon vertex is given in leading order (neglecting
loops) by the variational kernel $\gamma_4$ Eq.~\eqref{var3} determined in Sec.~\ref{subsec:xiphi}.
A form factor for the four-gluon vertex can be introduced along the same line of Eq.~\eqref{3gvff1}
\be\label{4gvff1}
f_{4A} \mathrel{\mathop:}= \frac{\Gamma^{(0)}_4 \circ \gamma_4}{\Gamma^{(0)}_4 \circ \Gamma^{(0)}_4} \, ,
\ee
where $\Gamma^{(0)}_4$ is the perturbative four-gluon vertex, which is obtained from
Eq.~\eqref{var3} by the replacements $\Omega(\vp)\to|\vp|$, $F(\vp)\to1/\vp^2$. We
consider here the form factor $f_{4A}$ at the symmetric point, where
\be\label{symmpt4gv}
\vp_1^2 = \ldots = \vp_4^2 = p^2 \, , \qquad \vp_i\cdot\vp_j = - \frac13 \, p^2 \quad (i\neq j) \, .
\ee
For this kinematic configuration the terms in Eq.~\eqref{var3} stemming from the Coulomb
interaction do not contribute to this vertex, and one finds for the form factor \eqref{4gvff1}
\be\label{4gvff2}
f_{4A}(p^2) = \frac{p}{\Omega(p^2)} \:
\frac{171-960 [g_0 + g(p^2)] + 8704 \, g_0 \, g(p^2)}
     {171-1920 \, g_0 + 8704 \, g_0^2}
\ee
where
\begin{subequations}\label{4gvff3}
\begin{align}
g(p^2) &= \left[ \frac{p}{2\Omega(p^2)+\Omega(\tfrac43 p^2)} \right]^{2} \label{4gvff3a} , \\
g_0 &= g(p^2)\Bigr\rvert_{\Omega(p^2)=p} = \tfrac38 (2 - \sqrt{3}) \, .  \label{4gvff3b}
\end{align}
\end{subequations}
The function $f_{4A}$ Eq.~\eqref{4gvff2} is shown in Fig.~\ref{fig:4gv}.
The function multiplying $p/\Omega(p^2)$ in Eq.~\eqref{4gvff2} is of $\mathcal{O}(1)$,
and the form factor at the symmetric point Eq.~\eqref{symmpt4gv} can be fairly well
represented by the gluonic dressing function $p/\Omega(p^2)$ alone.

In the above calculation of the four-gluon vertex we have neglected all loop diagrams,
in particular the ghost loop. From analogous investigations in Landau gauge \cite{KelFis08}
one may expect that the ghost loop dominates the IR behaviour of the four-gluon vertex
as it did for the three-gluon vertex. We will defer this issue to future research.


\section{\label{sec:summary}Summary and conclusions}
We have presented a general method to treat non-Gaussian wave functionals in the
Hamiltonian formulation of quantum field theory. By means of well-established
Dyson--Schwinger equation techniques, the equal-time Green functions and, in particular,
the expectation value of the Hamiltonian are expressed in terms of kernels occurring
in the exponent of the vacuum wave functional. These kernels are then determined by
the variational principle minimizing the vacuum energy density. The method was applied
to Yang--Mills theory in Coulomb gauge using a vacuum wave functional which contains
up to quartic terms in the exponent. In leading order (in the number of loops) the
cubic and quartic interaction kernels obtained from the variational principle are
reminiscent of the corresponding expressions obtained in leading order of a perturbative
solution of the Yang--Mills Schr\"odinger equation, except that the unperturbed gluon
propagators are replaced by the full ones. We have estimated these interaction kernels
$\gamma_{3,4}$ and the corresponding proper gluon vertex functions $\Gamma_{3,4}$
by using the gluon and ghost propagators found in the variational approach with
a Gaussian wave functional. The resulting three-gluon vertex compares
fairly well to the available lattice data obtained in $d=3$ Landau gauge. The (gap)
equation of motion obtained from the variation of the energy density with respect to
the gluon propagator contains the gluon loop, which was missed in previous variational
approaches due to the use of a Gaussian wave functional. We have shown that the gluon
loop gives a substantial contribution in the mid-momentum regime while leaving the IR
sector unchanged. Furthermore (together with the contribution from the non-Abelian
Coulomb interaction already fully included in previous studies \cite{FeuRei04}), it also provides
the correct asymptotic UV behaviour of the gluon propagator in accord with perturbation
theory \cite{CamReiWeb09}. The approach developed in the present paper allows a systematic
treatment of correlations in the Hamiltonian approach to interacting quantum field
theories (analogous to the so-called ``exponential $S$'' method in many-body physics)
and opens up a wide range of applications. In particular, it allows us to extend the
variational approach from pure Yang--Mills theory to full QCD. This will be subject
to future research.

\begin{acknowledgments}
The authors are grateful to P.\ Watson and J.\ M.\ Pawlowski for useful discussions
and for a critical reading of the manuscript. They also thank A.\ Maas for providing
the lattice data for the three-gluon vertex shown in Fig.~\ref{fig:3gv-lat}. This work
has been supported by the Deutsche Forschungsgemeinschaft (DFG) under contract
No.\ DFG-Re856-3 and by the Cusanuswerk--Bisch\"ofliche Studienf\"orderung.
\end{acknowledgments}


\end{document}